\documentclass[12pt,a4paper]{article}
\usepackage[pdftex,colorlinks=true]{hyperref}
\usepackage{amssymb,enumitem,booktabs,subfig,xcolor,todonotes,microtype,setspace,paralist,fullpage, makecell, longtable, bm, orcidlink, dsfont}
\usepackage[top=1.4in, bottom=1.4in, left=1.35in, right=1.35in]{geometry}
\usepackage{natbib}
\usepackage{url} 
\definecolor{darkblue}{rgb}{0,0,.6}
\hypersetup{citecolor=darkblue,linkcolor=darkblue,urlcolor=darkblue}
\usepackage{amsmath}
\usepackage{amsfonts}
\usepackage{epsfig}
\usepackage{graphics}
\usepackage{palatino,eulervm}
\usepackage{graphicx}
\usepackage{lscape}
\usepackage{rotating}
\usepackage[linewidth=1pt]{mdframed}
\allowdisplaybreaks[4]
\DeclareMathOperator*{\argmin}{arg\,min}

\newsavebox\CBox
\def\textBF#1{\sbox\CBox{#1}\resizebox{\wd\CBox}{\ht\CBox}{\textbf{#1}}}
\newcommand{\commHS}[1]{{\leavevmode\color{darkblue}#1}}

\providecommand{\U}[1]{\protect\rule{.1in}{.1in}}

\graphicspath{{plots/}}

\usepackage{amsthm,thmtools}
\usepackage{mathrsfs}

\makeatletter
\def\th@newremark{\th@remark\thm@headfont{\bfseries}}
\makeatletter
\theoremstyle{newremark}

\declaretheoremstyle[
  spaceabove=6pt, spacebelow=6pt,
  headfont=\bfseries,
  notefont=\mdseries, notebraces={(}{)},
bodyfont=\normalfont,
  postheadspace=0.5em,
]{mystyle}

\begin{document}

\title{\Large\bf Forecasting age distribution of life-table death counts via $\alpha$-transformation}
\author{\normalsize Han Lin Shang\orcidlink{0000-0003-1769-6430}\thanks{Corresponding address: Department of Actuarial Studies and Business Analytics, Macquarie University, Sydney, NSW 2068, Australia; Telephone number: +61(2) 9850 4689; Email: hanlin.shang@mq.edu.au}\\ \normalsize Department of Actuarial Studies and Business Analytics \\ \normalsize Macquarie University \\ \\
\normalsize Steven Haberman\orcidlink{0000-0003-2269-9759}\\ \normalsize Bayes Business School \\ \normalsize City St George's, University of London}
\date{\normalsize \today}

\maketitle

\begin{abstract}
We introduce a compositional power transformation, known as an $\alpha$-transformation, to model and forecast a time series of life-table death counts, possibly with zero counts observed at older ages. As a generalisation of the isometric log-ratio transformation (i.e., $\alpha=0$), the $\alpha$ transformation relies on the tuning parameter $\alpha$, which can be determined in a data-driven manner. Using the Australian age-specific period life-table death counts from 1921 to 2020, the $\alpha$ transformation can produce more accurate short-term point and interval forecasts than the log-ratio transformation. The improved forecast accuracy of life-table death counts is of great importance to demographers and government planners for estimating survival probabilities and life expectancy and actuaries for determining annuity prices and reserves for various initial ages and maturity terms. 

\vspace{.1in}
\noindent \textit{Keywords}: compositional data analysis; centre log-ratio; isometric log-ratio; principal component analysis; functional time-series forecasting
\end{abstract}

\newpage

\section{Introduction}\label{sec:1}

Actuaries and statistical demographers have been interested in developing models for mortality forecasting \citep[see][for comprehensive reviews]{Booth06, BT08, BCB22}. From an actuarial perspective, mortality forecasts are important to manage adverse financial effects of mortality improvements over time on life or fixed-term annuities and sustainable pension system \citep{Pollard87}. From a demographic perspective, mortality forecasts are an essential part of forward planning for national health and aged care systems and are an important component of population forecasts \citep[see, e.g.,][among others]{SSB+16, HZS21}.

In demography, three functions related to mortality are widely studied: central mortality rates, survival probabilities, and life-table death counts. Several scholars have proposed new approaches for forecasting age-specific mortality rates using statistical models \citep[see, e.g.,][for reviews]{Booth06, BT08, CBD08, SBH11}. Arguably, the most famous model is the \cite{LC92} model. Let $m_{x,t}$ be log central mortality rates for age $x$ in year $t$. The Lee-Carter model is given by
\begin{equation}
m_{x,t} = a_x + b_x \kappa_t + \epsilon_{x,t}, \quad t=1,2,\dots,n, \; x=0,1,\dots, p, \label{eq:LC}
\end{equation}
where $a_x$ is the age pattern of the log central mortality rates averaged across years; $b_x$ is the first principal component capturing relative change in the log central mortality rate at each age $x$ with a constraint $\sum^p_{x=0}b_x=1$; $\kappa_t$ is the first set of principal component scores measuring general level of the log central mortality rate at year~$t$ with a constraint $\sum^n_{t=1}\kappa_t = 0$; and $\epsilon_{x,t}$ is the model residual at age~$x$ and year~$t$. Both constraints ensure the parameter identifiability in the Lee-Carter model. The principal component analysis is performed on the mean corrected matrix of log mortality rates. The principal component scores $\kappa_t$ are extrapolated by a random walk with drift method, from which point forecasts are obtained by~\eqref{eq:LC} with the fixed $a_x$ and $b_x$. In demography, there exist several extensions and modifications of the Lee-Carter model \citep[see, e.g.,][for review]{BT08, SBH11, BCB22}.

Apart from modelling central mortality rates, one can also model a redistribution of life-table death counts over time \citep[see, e.g.,][]{PLC19, BKC20}. For instance, in many developed countries, deaths at younger ages have been gradually shifting toward older ages. The shift of the distribution symbolises longevity risk, which is a major issue for life insurers and pension funds, especially in the selling and risk management of annuity products \citep[see][for a discussion]{BDV02, DDG07}. In addition to providing an informative description of the mortality experience of a population, the life-table death counts yield readily available information on ``central longevity indicators', i.e., mean, median, and mode age at death \citep{CanudasRomo10}, and lifespan variability \citep{Robine01, VZV11, VC13, VMM14}. For many developed countries, a decrease in variability over time may be observed through the standard deviation of life-table ages at death, Gini coefficient \citep{WH99, VC13, SHX22} or Drewnowski's index \citep{ABB+22}. Life-table death counts provide important insights into longevity risk and lifespan variability that cannot be easily quantified from either central mortality rates or survival probabilities.

Life-table death counts relate strongly with the probability density function. For each year, the life-table death counts are non-negative and sum to a radix of $10^5$. Because of the two constraints, the age-specific life-table death counts can be viewed as compositional data. The sample space of compositional data is a simplex
\[
S^{D-1} = \left\{(d_1, d_2, \dots,d_D)^{\top}, \quad d_x>0, \quad \sum^D_{x=1}d_x = c\right\},
\]
where $d_x$ denotes life-table death count for age $x$, $D$-part compositional data are mapped from the simplex into a $(D-1)$-dimensional real space, $c$ is a fixed constant, and $^{\top}$ denotes vector transpose.

Compositional data are defined as a random vector of $D$ compositions $[d_1, d_2,\dots,d_D]$ with non-negative values whose sum is a given constant, typically one (portions), 100 (\%) and $10^6$ for parts per million in geochemical compositions. These data arise in many scientific fields, such as geology (geochemical compositions), economics (income/expenditure distribution), medicine (body composition), food industry (goods composition), chemistry (chemical composition), ecology (abundance of different species), and demography. In statistics, \cite{SDG+15} apply compositional data analysis (CoDa) to study the concentration of chemical compositions in sediment or rock samples. In economics, \cite{SW17} apply CoDa to analyse the total weekly household expenditure on food and housing costs. \cite{KMP+19} apply CoDa to model and forecast density functions of financial assets. In demography, \cite{Oeppen08}, \cite{BCO+17}, \cite{SH20}, and \cite{SHX22} introduce a principal component approach to model and forecast life-table death counts within a compositional data analytic framework.

In the CoDa, \cite{AS80} and \cite{Aitchison1982, Aitchison1986} transform the compositional data from the simplex to Euclidean space using a centre log-ratio (clr) transformation:
\[
\bm{s} = \{s_x\}_{x=1,\dots,D} = \ln\left\{\frac{d_x}{g(d)}\right\}_{x=1,\dots,D},
\]
where $g(d) = (\prod^D_{x=1}d_x)^{1/D}$. Extending from the clr transformation, \cite{EPM+03} propose the isometric log-ratio (ilr) transformation, and it has been promoted as the more theoretically correct method in CoDa \citep{GG19}. The ilr transformation is defined as:
\[
\bm{z} = \bm{H}\bm{s}
\]
where $\bm{H}$ is a $(D-1)\times D$ Helmert sub-matrix \citep[see, e.g.,][]{Lancaster65, DM98}. In geographic mapping, Helmert sub-matrices are used to describe a coordinate transformation. The Helmert sub-matrix of order $D$ is a square matrix defined as
\[
\bm{\mathcal{H}}_D = \begin{bmatrix}
\frac{1}{\sqrt{D}} & \frac{1}{\sqrt{D}} & \frac{1}{\sqrt{D}} & \cdots & \frac{1}{\sqrt{D}} \\
\frac{1}{\sqrt{2}} & -\frac{1}{\sqrt{2}} & 0 & \cdots & 0 \\
\frac{1}{\sqrt{6}} & \frac{1}{\sqrt{6}} & -\frac{2}{\sqrt{6}} & \cdots & 0 \\
\vdots & \vdots & \vdots & \ddots & \vdots \\ 
\frac{1}{\sqrt{D(D-1)}} & \frac{1}{\sqrt{D(D-1)}} & \frac{1}{\sqrt{D(D-1)}} & \cdots & -\frac{(D-1)}{\sqrt{D(D-1)}} \\
\end{bmatrix}.
\]
$\bm{H}=(\bm{\mathcal{H}}_D)_{(D-1)\times D}$ is introduced to remove any redundant dimension presented due to the compositional constraint. The $\bm{H}$ matrix is an orthogonal matrix without the first row.

An issue with any log-ratio transformation, including the clr and ilr, is that there is no guarantee that the transformed data are multivariate normal distributed \citep{TS20}. This distributional assumption plays a crucial role in the validity of the log-ratio analysis \citep{AS80}. Another issue is that the log-ratio transformation cannot be applied to data containing zeros. The presence of zero counts makes the log transformation invalid. In statistical demography, there exist some missing-value imputation methods for handling zeros, but these methods are often ad-hoc by omitting, aggregating, or adding small arbitrary values to zero values \citep[see, e.g.,][]{FFM00, MBP03}. This is not ideal, however, \cite{Greenacre21} compared four different algorithms to substitute zeros and showed that the resulting conclusions can be strongly sensitive to the method of zero substitution. In a similar spirit to the Box-Cox transformation, we present a power transformation, known as $\alpha$ transformation, which can handle zero counts.

The paper is organised as follows. In Section~\ref{sec:3}, we show that the log-ratio analysis can be seen as a special case of $\alpha$ transformation, where the optimal value of $\alpha$ can be determined in a data-driven way, and this is discussed in Section~\ref{sec:3.3}. Via the $\alpha$ transformation, the functional time-series forecasting method can be applied to model and forecast the unconstrained data. We introduce the Australian age-specific life-table death counts in Section~\ref{sec:2}, which we use to illustrate the methodology. In Section~\ref{sec:4}, we describe the model fitting and forecasting and introduce the measures of point forecast error in Section~\ref{sec:4.3}, which we use in Section~\ref{sec:4.4} to evaluate and compare the point forecast accuracy. Using the coverage probability difference and mean interval score in Section~\ref{sec:4.5}, we evaluate and compare the interval forecast accuracy in Section~\ref{sec:4.6}. The conclusion is presented in Section~\ref{sec:5}, along with some ideas on how the methodology may be further extended.

\section{Age- and sex-specific life-table death counts}\label{sec:2}

We study Australian age- and sex-specific life-table death counts from 1921 to 2020, obtained from the \cite{HMD2023}. The life-table radix is fixed at $10^5$ at age 0 and gradually decreases to 0 at age 110+ for every calendar year. Due to rounding, there may be zero counts for older ages at some years. One potential remedy is to use the probability of dying (i.e., $q_x$) and the life-table radix to recalculate our estimated age-specific death counts \citep[see, e.g.,][]{SH20}. The estimated death counts are more detailed and smoother than those reported in the \cite{HMD2023}. 

\begin{figure}[!htb]
\centering
\includegraphics[width=8.7cm]{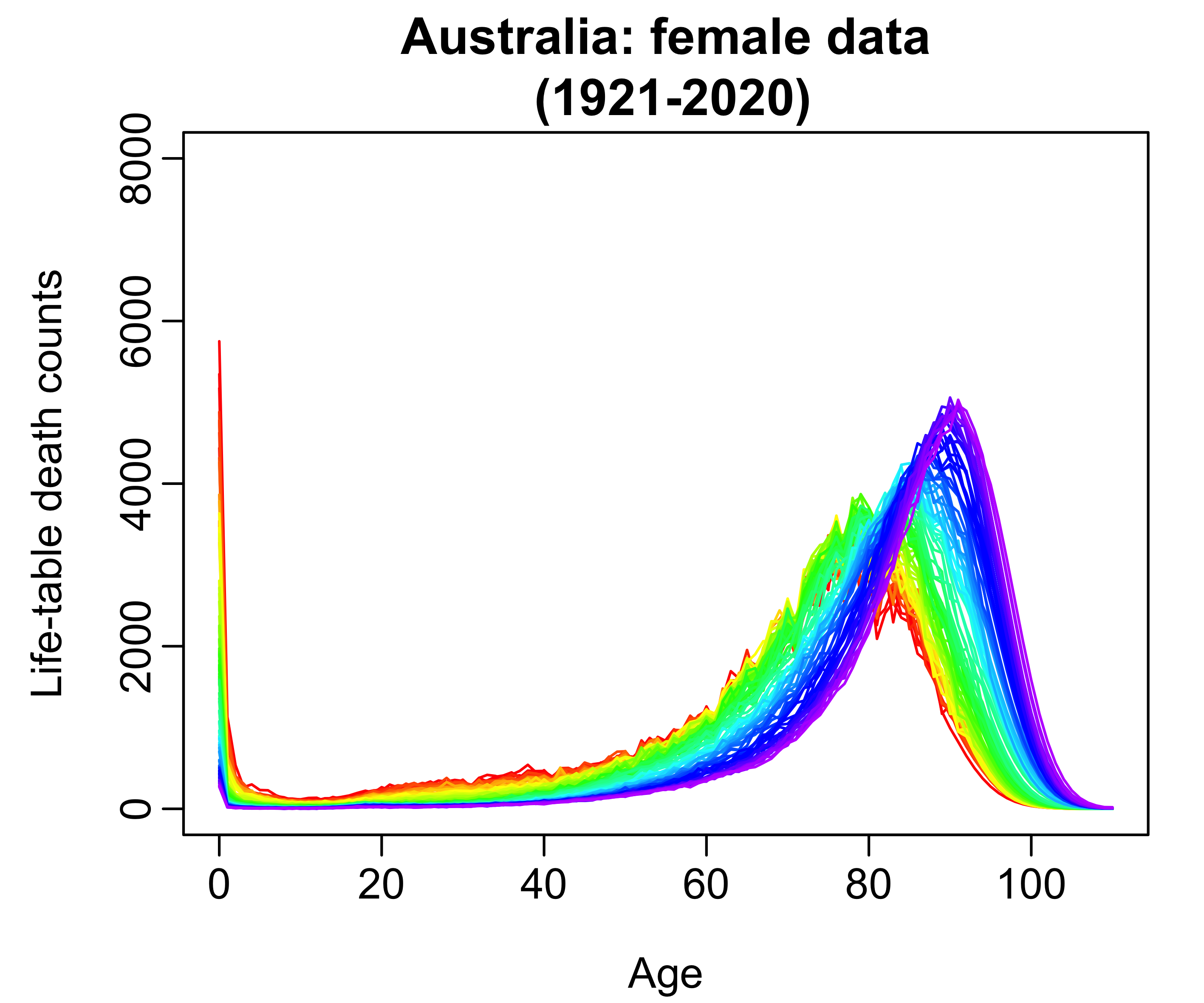}
\quad
\includegraphics[width=8.7cm]{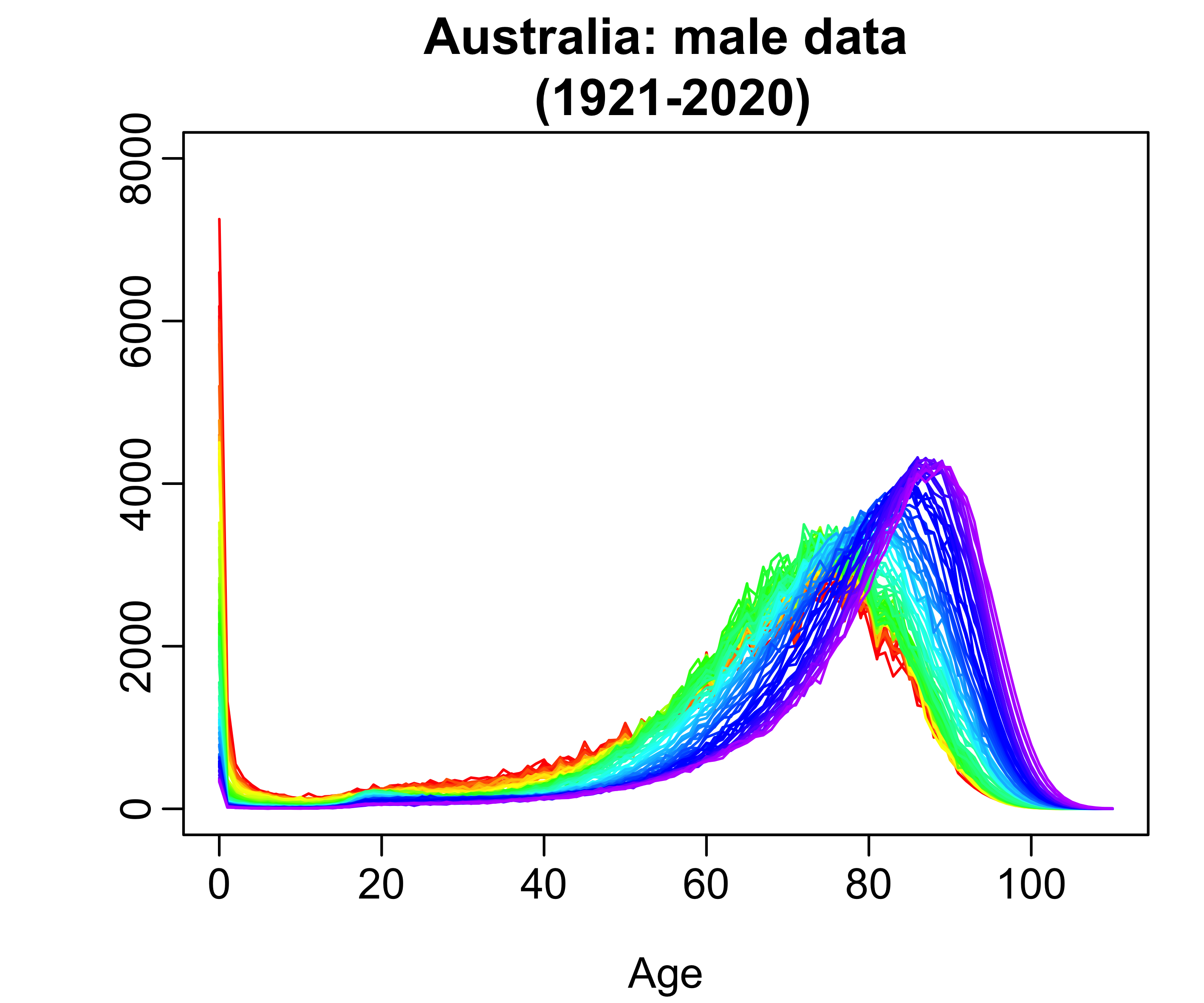}
\caption{\small Rainbow plots of age-specific life-table death count from 1921 to 2020 in a single-year group. The oldest years are shown in red, with the most recent in violet. Curves are ordered chronologically according to the colours of the rainbow.}\label{fig:1}
\end{figure}

In Figure~\ref{fig:1}, we present rainbow plots of the female and male age-specific life-table death counts in Australia for a single year of age groups. Both panels of Figure~\ref{fig:1} demonstrate a decreasing trend in infant mortality and a negatively skewed distribution for the life-table death counts, where the peaks shift to older ages for females and males. The gradual change in distribution confirms the longevity risk experienced in recent years.

\section{The $\alpha$ transformation}\label{sec:3}

One common approach in compositional data analysis involves transforming compositional data to address non-normality and heteroscedasticity, with the $\alpha$ transformation being one such technique. Akin to Box-Cox transformation, the $\alpha$ transformation of a compositional vector $\bm{d}_t \in \mathbb{S}^D$ is defined by
\begin{equation}
\bm{z}_t^{\alpha} = \bm{H}\cdot \left(\frac{D\bm{d}_{t}^{\alpha}-\mathds{1}_D}{\alpha}\right)\label{eq:alpha-transform}
\end{equation}
where $\alpha\in [0,1]$ and $\bm{d}_{t}^{\alpha}$ is given as
\[
\bm{d}_{t}^{\alpha} = \left(\frac{d_{t,1}^{\alpha}}{\sum^D_{j=1}d_{t,j}^{\alpha}}, \dots, \frac{d_{t,D}^{\alpha}}{\sum^D_{j=1}d_{t,j}^{\alpha}}\right)^{\top},
\]
and $\mathds{1}_D$ is the $D$-dimensional vector of ones, and $\bm{H}$ is $(D-1)\times D$ Helmert matrix. The compositional power transformation is invertible:
\begin{equation}
\bm{\nu}_t^{\alpha}=D\bm{d}_{t}^{\alpha}=\alpha\times\bm{H}^{\top} \bm{z}_t^{\alpha} +\mathds{1}_{D},\label{eq:alfa_inv}
\end{equation}
where 
\[
\bm{d}_t=\left(\frac{v_{t,1}^{1/\alpha}}{\sum^D_{j=1}v_{t,j}^{1/\alpha}},\dots, \frac{v_{t,D}^{1/\alpha}}{\sum^D_{j=1}v_{t,j}^{1/\alpha}}\right).
\]
As noted in \cite{TPW16}, when $\alpha=0$, it corresponds to the isometric log-ratio transformation, whereas $\alpha=1$, it corresponds to Euclidean data analysis (eda) ignoring the compositional constraint \citep[see, e.g.,][]{Baxter95, Baxter01, BBC+05}. When $\alpha = 0$ and one does not pre-multiply Helmert matrix $\bm{H}$, the formulation reduces to the clr.

\subsection{Modelling and forecasting a time series of density-valued curves}\label{sec:3.1}

Through the $\alpha$ transformation, we obtain a time series of unconstrained matrix $\bm{z}^{\alpha} = (\bm{z}_1^{\alpha}, \dots, \bm{z}_n^{\alpha})$ where $\bm{z}_1^{\alpha} = (z_{1,1}^{\alpha},\dots,z_{1,D}^{\alpha})^{\top}$. We apply a principal component analysis to decompose $\bm{z}^{\alpha}$. Under a finite second moment, we compute the sample covariance of $\bm{z}^{\alpha}$. Via Mercer's lemma, the sample covariance, denoted by $\widehat{\bm{C}}$, can be estimated by
\[
\widehat{\bm{C}} = \sum_{k=1}^{\infty} \widehat{\lambda}_k \widehat{\bm{\phi}}_{k} \widehat{\bm{\phi}}_{k},
\]
where $\widehat{\lambda}_1> \widehat{\lambda}_2> \dots$ are non-increasing eigenvalues. By Karhunen-Lo\`{e}ve expansion, for a given year $t$,
\[
z_t^{\alpha}=\overline{z}^{\alpha} + \sum^K_{k=1}\widehat{\beta}_{t,k}\widehat{\bm{\phi}}_k + e_t,
\]
where $\overline{z}^{\alpha}=\frac{1}{n}\sum^n_{t=1}z_t^{\alpha}$ is the mean term, $\widehat{\beta}_{t,k}$ denotes the $k$\textsuperscript{th} set of the estimated principal component score for time $t$, $\widehat{\bm{\phi}}_k$ denotes the $k$\textsuperscript{th} set of the estimated principal component, and $e_t$ denotes the model error term encompassing the remaining principal component pairs. 

To determine the optimal number of principal components, we select $K$ via an eigenvalue ratio criterion based on \cite{LRS20}. 
\begin{equation}
K = \argmin_{1\leq k\leq K_{\max}}\left\{\frac{\widehat{\lambda}_{k+1}}{\widehat{\lambda}_k}\times \mathds{1}\Big\{\frac{\widehat{\lambda}_k}{\widehat{\lambda}_1}\geq \theta\Big\}+\mathds{1}\Big\{\frac{\widehat{\lambda}_k}{\widehat{\lambda}_1}<\theta\Big\}\right\},\label{eq:eigenvalue_ratio}
\end{equation}
where $\mathds{1}\{\cdot\}$ denotes the binary indicator function, $\theta = \frac{1}{\ln [\max(\widehat{\lambda}_1,n)]}$ is a small positive value, and instead of searching through $n$ sets of principal component pairs, we restrict our searching range by setting $K_{\max} = \#\{k|\widehat{\lambda}_k\geq \frac{1}{n}\sum^n_{k=1}\widehat{\lambda}_k, k\geq 1\}$ \citep[see also][]{LRS20}. Equation~\eqref{eq:eigenvalue_ratio} can be viewed as a ridge-type eigenvalue ratio criterion. The functional time-series forecasting method is robust to over-estimating $K$, but under-estimating $K$ can result in inferior accuracy. As in \cite{HBY13}, we also consider $K=6$.

Having determined the optimal number of principal components $K$, we apply an autoregressive integrated moving average (ARIMA) model to obtain the $h$-step-ahead forecast. Conditional on the historical data and estimated mean and principal components, the $h$-step-ahead forecast of $\bm{z}_{n+h}^{\alpha}$ can be expressed as
\[
\bm{\widehat{z}}_{n+h|n}^{\alpha} = E[z_{n+h}^{\alpha}|\bm{\widehat{\Phi}}, \bm{z}^{\alpha}] = \overline{z}^{\alpha} + \sum^K_{k=1}\widehat{\beta}_{n+h|n,k}\widehat{\bm{\phi}}_k,
\]
where $\bm{\widehat{\Phi}} = (\widehat{\bm{\phi}}_1,\dots,\widehat{\bm{\phi}}_K)^{\top}$, and $\widehat{\beta}_{n+h|n,k}$ denotes the $h$-step-ahead forecast of the $k$\textsuperscript{th} principal component scores. By taking the inverse $\alpha$-transformation in~\eqref{eq:alfa_inv}, we obtain the $h$-step-ahead forecast, denoted by $\bm{\widehat{d}}_{n+h|n}=(\widehat{d}_{n+h|n,1},\dots,\widehat{d}_{n+h|n,D})$.

We use the automatic algorithm of \cite{HK08} to choose the optimal autoregressive order $p$, moving average order $q$, and difference order $d$ based on the optimal AIC with a correction for a small finite sample size \citep[see, e.g.,][]{HT93}. After identifying the optimal ARIMA model, the maximum likelihood method can be used to estimate the regression coefficients associated with the ARIMA model.

\subsection{Construction of pointwise prediction intervals for the $\alpha$ transformation}\label{sec:3.2}

Following \cite{HS09}, we incorporate two sources of uncertainty: truncation errors in the principal component decomposition and forecast errors in the forecast principal component scores. To obtain bootstrap forecasts of the scores, we assess the variability between the forecast and observed principal component scores. With a univariate time-series model, we can obtain multi-step-ahead forecasts for the scores, $\{\widehat{\beta}_{1,k}, \widehat{\beta}_{2,k},\dots,\widehat{\beta}_{n,k}\}$ for $k=1, 2, \dots,K$. Let the $h$-step-ahead forecast errors be given by $\zeta_{t,h,k} = \widehat{\beta}_{t,k} - \widehat{\beta}_{t|t-h,k}$ for $t=h+1,\dots,n$. These can then be sampled with replacement to give a bootstrap sample of $\beta_{n+h,k}$:
\[
\widehat{\beta}_{n+h|n,k}^{(b)} = \widehat{\beta}_{n+h|n,k} + \widehat{\zeta}_{*,h,k}^{(b)},\quad b=1,\dots,B,
\]
where $B=1,000$ symbolises the number of bootstrap replications and $\widehat{\zeta}_{*,h,k}^{(b)}$ are sampled with replacement from $\widehat{\zeta}_{t,h,k}$.

Assuming the first $K$ principal components approximate the data relatively well, the model residual should contribute nothing but random noise. Consequently, we can bootstrap the model residuals by sampling with replacement from $\{e_1, e_2,\dots,e_n\}$. 

Addinging two components of variability, we obtain $B$ variants for $\widehat{z}_{n+h|n}^{\alpha,(b)}$:
\[
\widehat{z}_{n+h|n}^{\alpha,(b)} = \overline{z}^{\alpha} + \sum^K_{k=1}\widehat{\beta}_{n+h|n,k}^{(b)}\widehat{\bm{\phi}}_k + e_{n+h}^{(b)},
\]
where $\widehat{\beta}_{n+h|n,k}^{(b)}$ denotes the forecast of bootstrapped principal component scores. With the bootstrapped $\{\widehat{z}_{n+h|n}^{\alpha,(1)}, \widehat{z}_{n+h|n}^{\alpha,(2)}, \dots, \widehat{z}_{n+h|n}^{\alpha,(B)}\}$, we take the inverse $\alpha$ transformation to obtain bootstrap forecasts for age distribution of death counts. At the $100(1-\gamma)\%$ nominal coverage probability, the pointwise prediction intervals are obtained by $\gamma/2$ and $1-\gamma/2$ quantiles based on $\{\widehat{\bm{d}}_{n+h|n}^{(1)}, \widehat{\bm{d}}_{n+h|n}^{(2)}, \dots, \widehat{\bm{d}}_{n+h|n}^{(B)}\}$, where $\gamma$ represents a level of significance, customarily $\gamma=0.2$ or 0.05.

\subsection{Selection of $\alpha$ parameter}\label{sec:3.3}

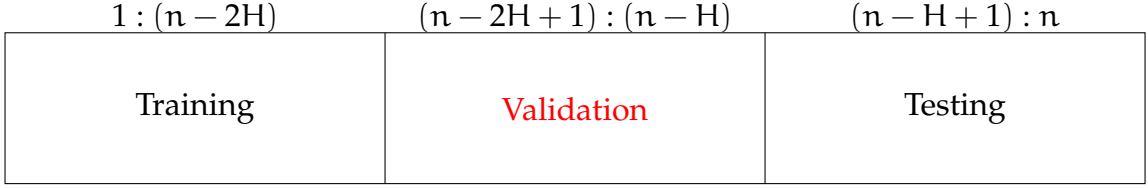
\begin{figure}[!htb]
\begin{center}
\begin{tikzpicture}
\draw (0,0) rectangle (15,2);
\draw (5,2) -- (5,0);
\draw (10,2) -- (10,0);
\draw (2.5,1) node {Training};
\draw (7.5,1) node[red] {Validation};
\draw (12.5,1) node {Testing};
\draw(2.5,2.2) node{$1:(n-2H)$};
\draw(7.5,2.2) node{$(n-2H+1):(n-H)$};
\draw(12.5,2.2) node{$(n-H+1):n$};
\end{tikzpicture}
\end{center}
\caption{\small Illustration of the cross-validation method. A model is constructed using data in the training set to forecast data in the validation set. The model's predictive ability is evaluated based on either point or interval forecast error. The optimal value of $\alpha$ is determined based on the minimal point or interval forecast error in the validation set. For sensitivity analysis, we consider $H=10$ or 20.}\label{fig:2}
\end{figure}

We split the data with a sample size of $n$ into a training sample and a testing set. Within the training sample, we further divide it into a training set and a validation set. With a maximum forecast horizon $H$, the testing set contains the elements $(n-H+1):n$, the validation set contains $(n-2H+1):(n-H)$, and the initial training set contains $1:(n-2H)$. To select the weight parameter, we aim to find $\alpha$ that minimises a point forecast error in Section~\ref{sec:4.3} and an interval forecast error in Section~\ref{sec:4.5}. In Figure~\ref{fig:2}, we present a visual display of the sample splitting.

\newpage

\begin{figure}[!htb]
\centering
\subfloat[$\alpha=0.3544$]
{\includegraphics[width=6.05cm]{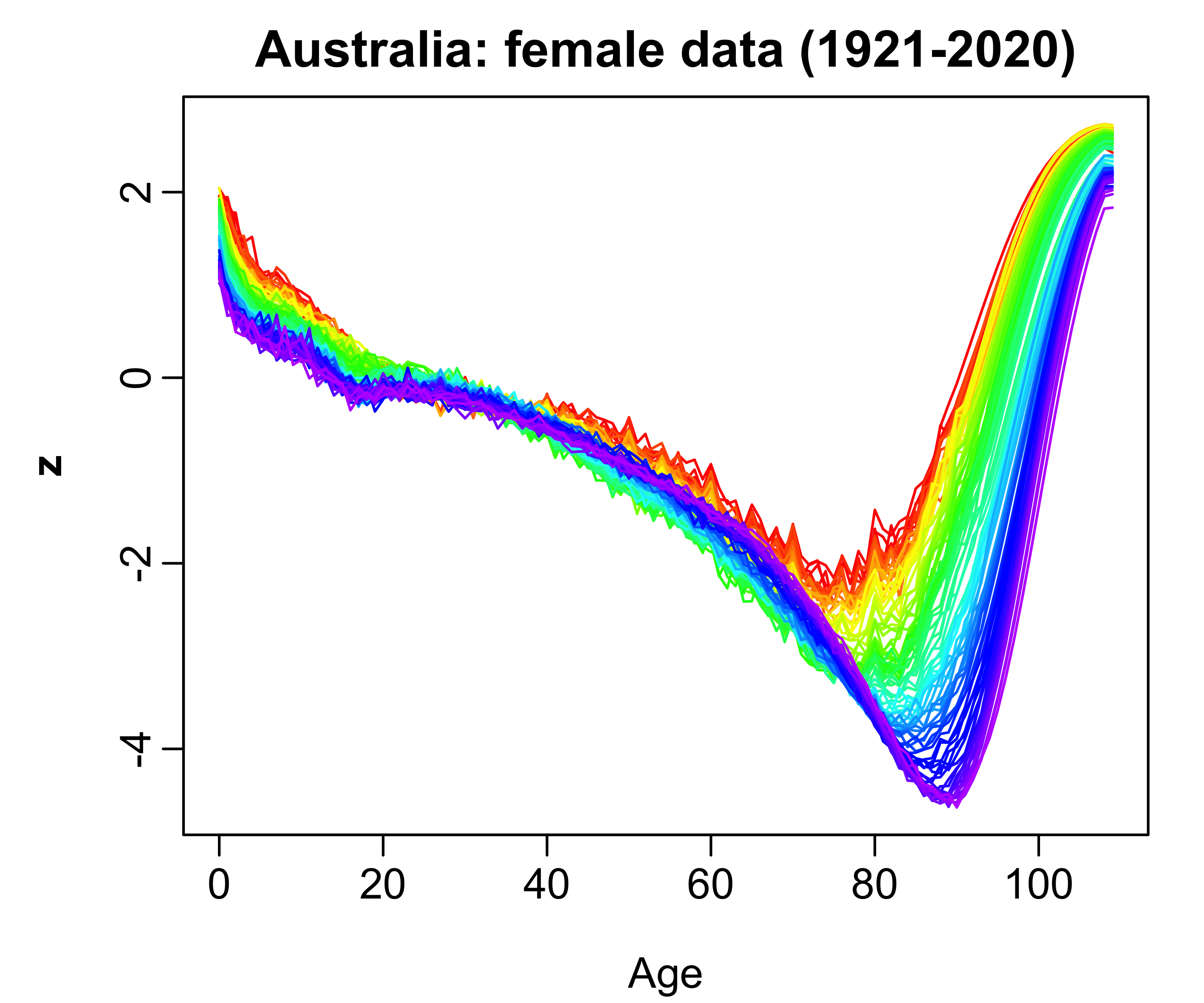}\label{fig:3a}}
\subfloat[$\alpha=0$]
{\includegraphics[width=6.05cm]{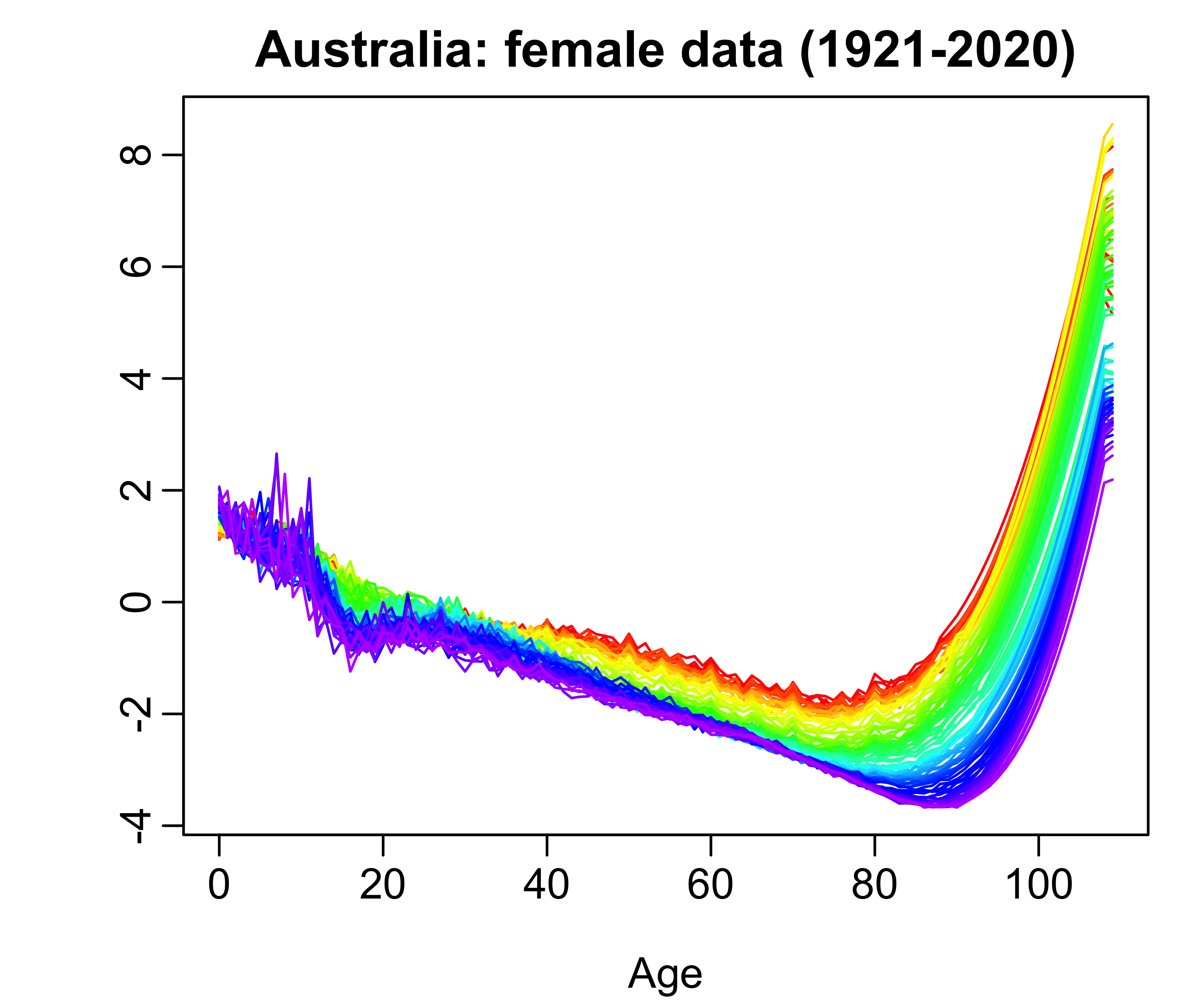}\label{fig:3b}}
\subfloat[$\alpha=1$]
{\includegraphics[width=6.05cm]{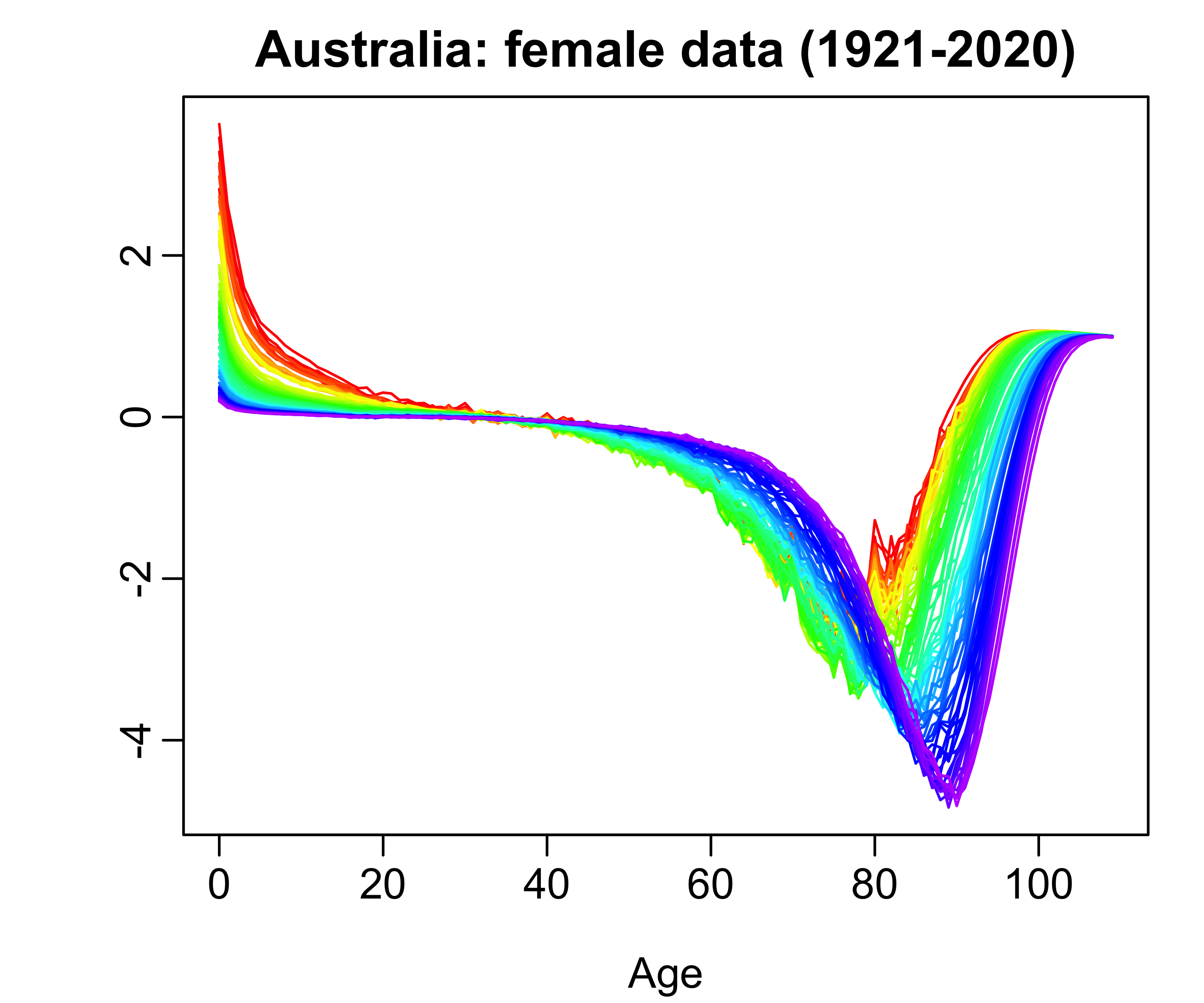}\label{fig:3c}}
\caption{\small The $\alpha$-transformed data of the Australian female life-table death counts. Based on the Kullback-Leibler divergence described in Section~\ref{sec:4.3}, $\alpha=0.3544$ is optimal. When $\alpha=0$, it corresponds to the ilr method; when $\alpha=1$, it corresponds to the eda method.}
\end{figure}

The forecast accuracy of $\alpha$ transformation depends on the optimal selection of the $\alpha$ value for a given forecast horizon. In Figure~\ref{fig:3a}, we present the $\alpha$-transformed data of female life-table death counts, where $\alpha = 0.3544$ is selected based on the Kullback-Leibler divergence for $H=10$. For comparison, we also show the transformed data where $\alpha = 0$ and 1. As one alters the value of $\alpha$, it impacts the shape of the transformed time series of densities.

\section{Model fitting and forecasting}\label{sec:4}

\begin{figure}[!htb]
\centering
\subfloat[Unconstrained female data]
{\includegraphics[width=6cm]{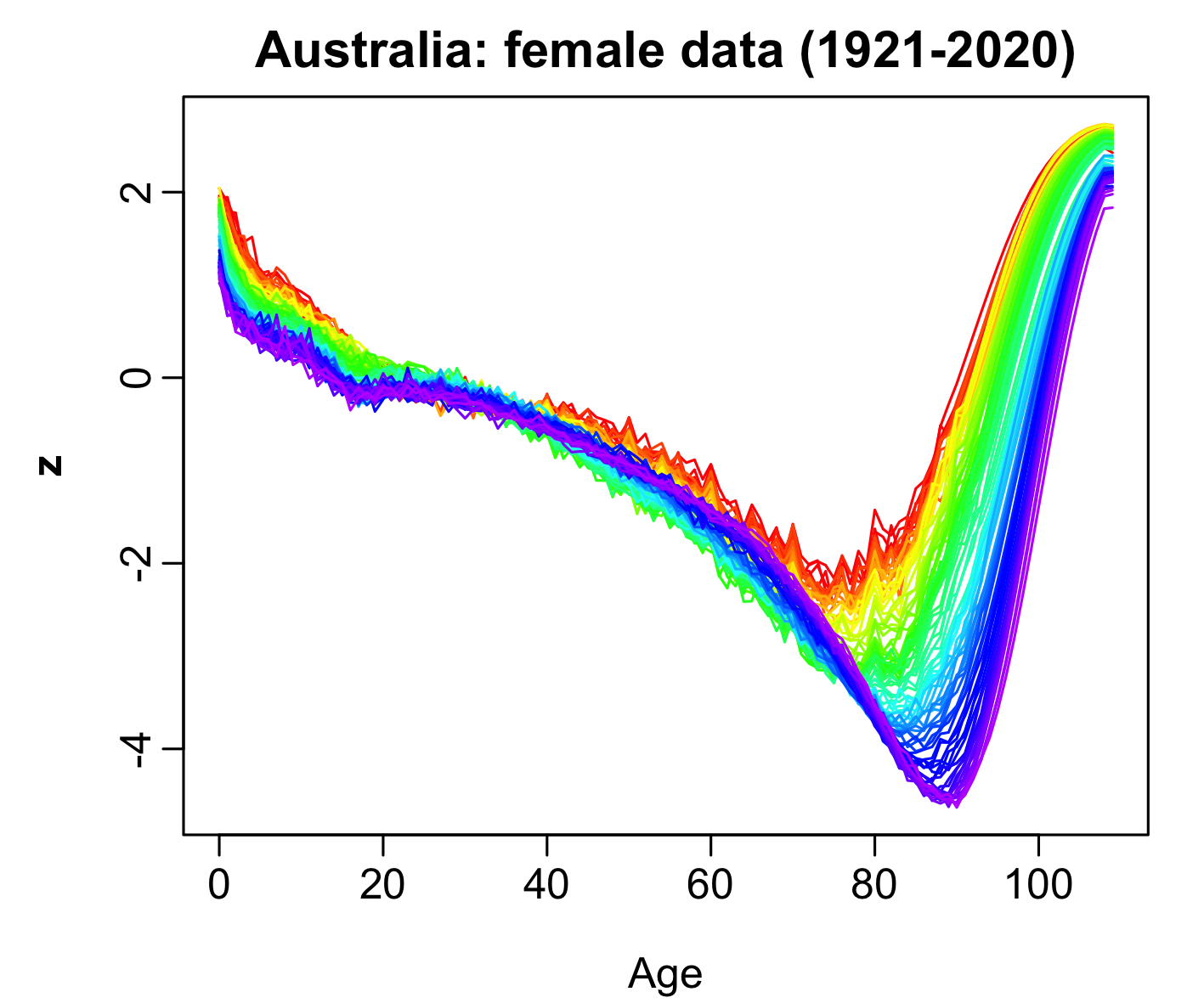}}
\subfloat[1$\textsuperscript{st}$ set of principal component]
{\includegraphics[width=6cm]{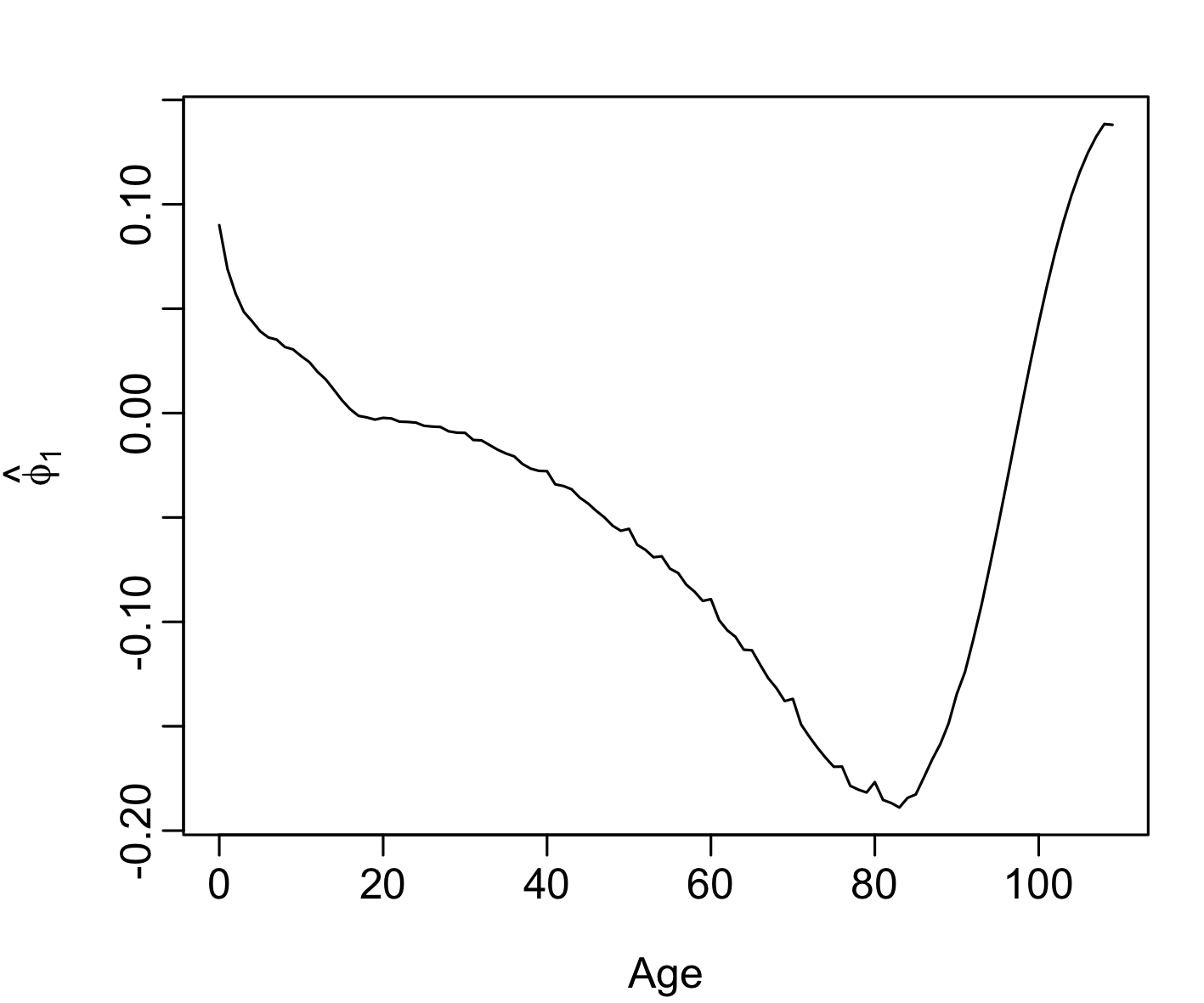}}
\subfloat[2$\textsuperscript{nd}$ set of principal component]
{\includegraphics[width=6cm]{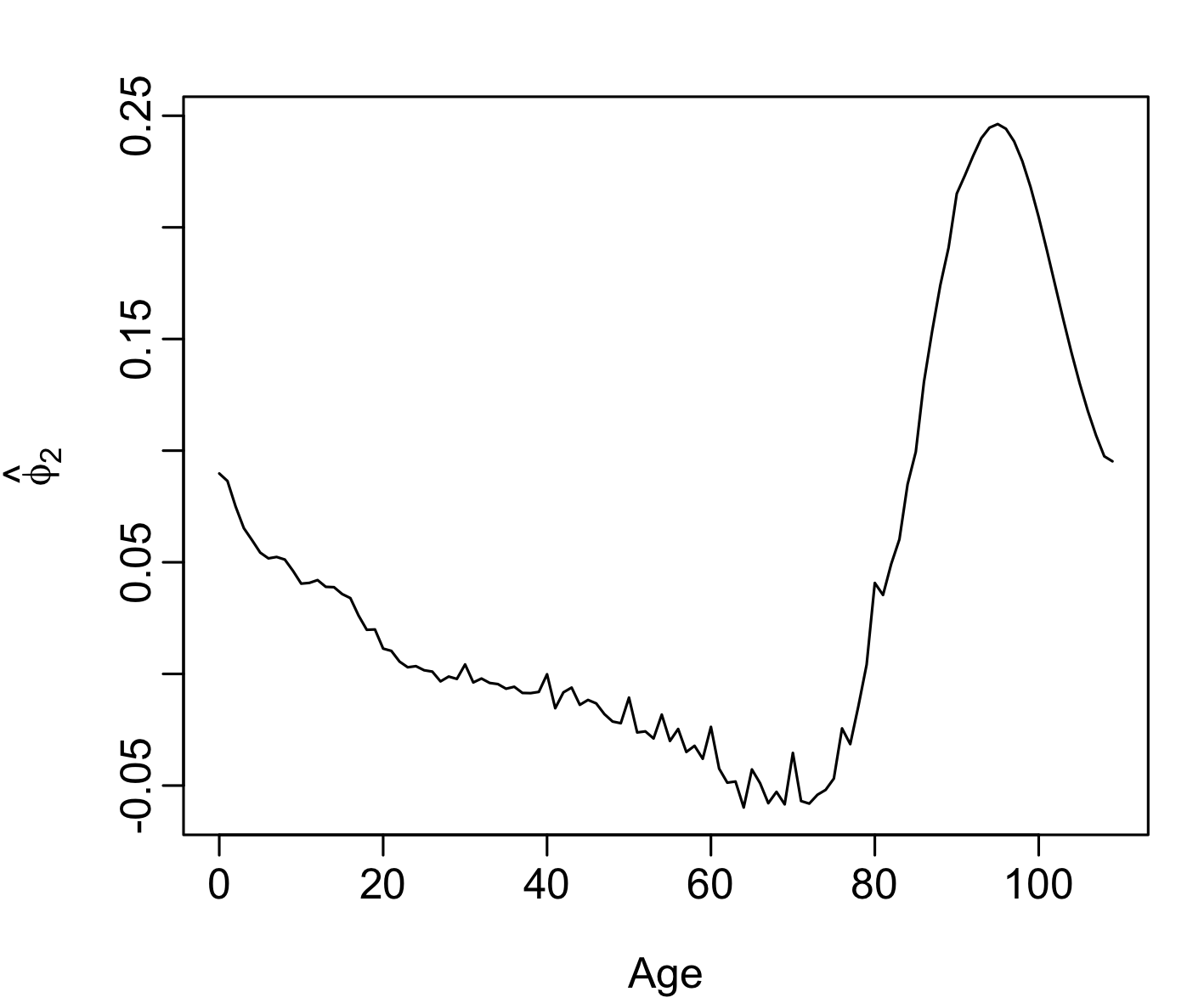}}
\\
\subfloat[Forecast life-table death counts]
{\includegraphics[width=6cm]{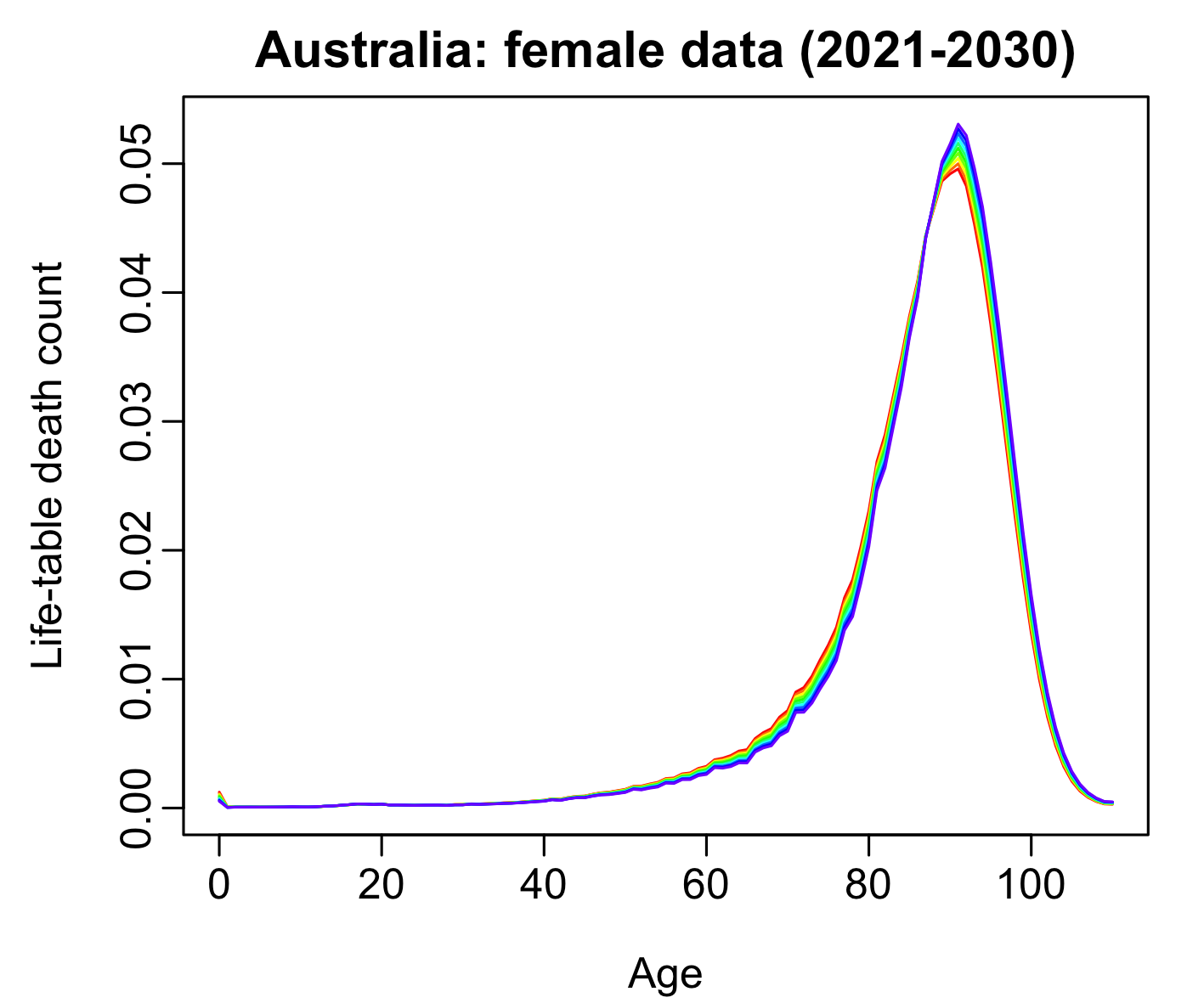}\label{fig:3d}}
\subfloat[Forecast 1$\textsuperscript{st}$ set of scores]
{\includegraphics[width=6cm]{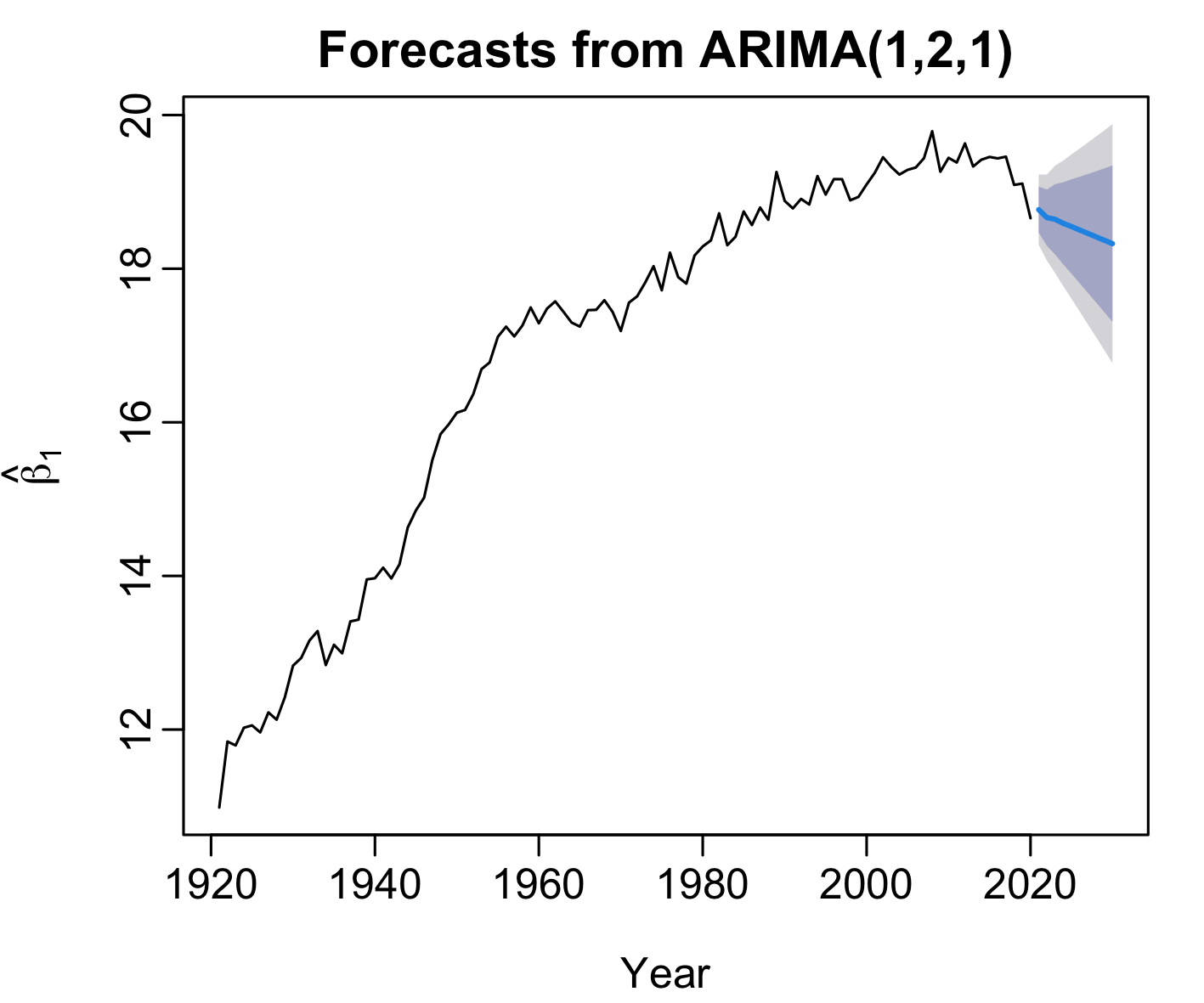}}
\subfloat[Forecast 2$\textsuperscript{nd}$ set of scores]
{\includegraphics[width=6cm]{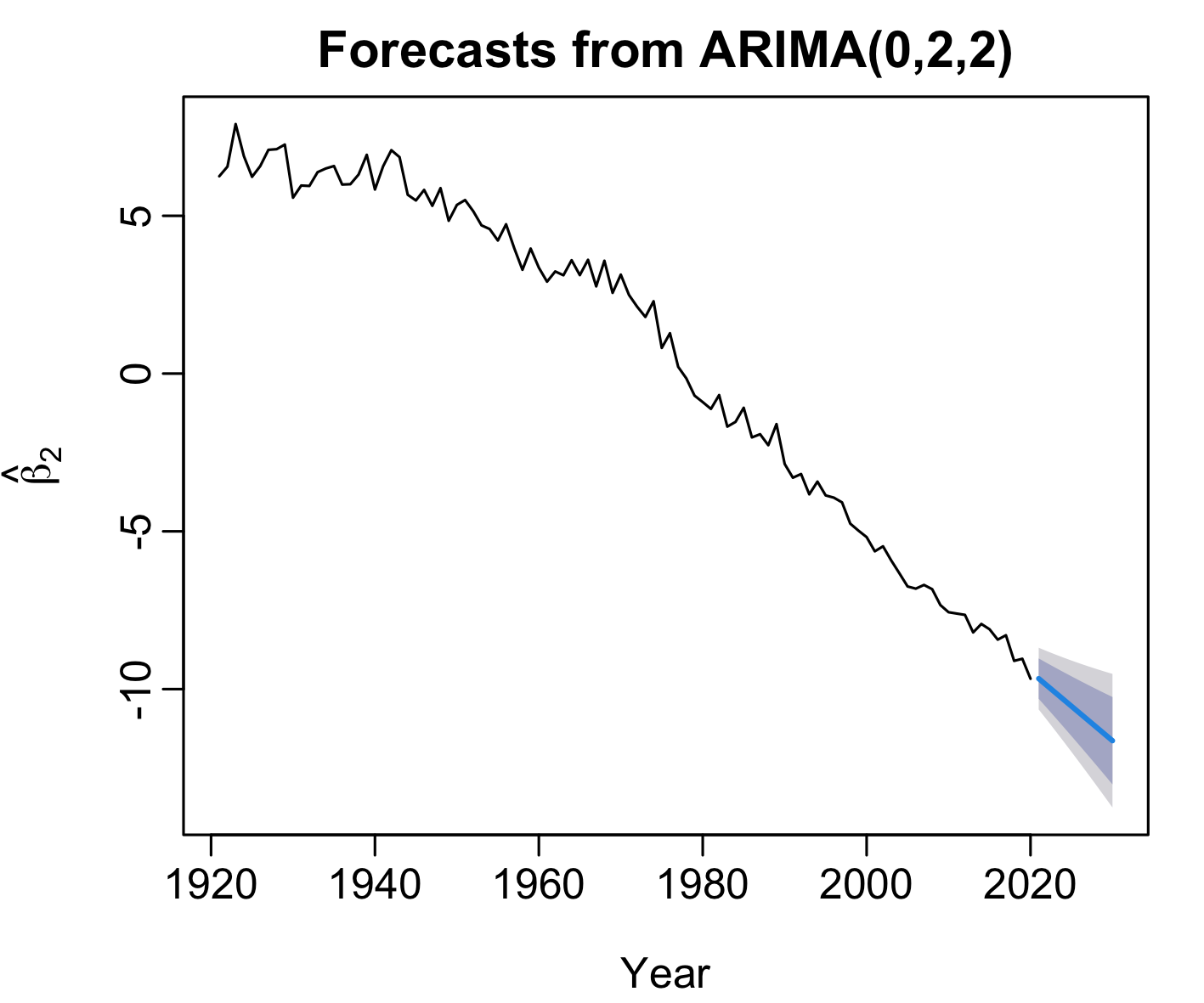}}
\caption{\small Elements of the $\alpha$-transformation for analysing the female age-specific life-table death counts in Australia. As determined by the eigenvalue criterion, we present the first two principal components and their associated principal component scores.}\label{fig:3}
\end{figure}

Based on the historical death counts from 1921 to 2020 (i.e., 100 observations), we apply the eigenvalue ratio criterion in~\eqref{eq:eigenvalue_ratio} to estimate the number of retained principal components, which is $K=2$. For modelling each set of the scores, we apply the automatic algorithm of \cite{HK08} to select the optimal orders of the ARIMA model shown in the second row of Figure~\ref{fig:3}. From the forecast principal component scores, one can observe an increasing trend indicating a continuation of the recent trend, i.e., compression of life-table death counts \citep[see also][]{SH20}. By multiplying the forecast principal component scores with the estimated functional principal components and adding the mean function, we produce 10-steps-ahead forecasts of life-table death counts between 2021 and 2030, shown in Figure~\ref{fig:3d}. 

\begin{figure}[!htb]
\centering
\subfloat[Unconstrained male data]
{\includegraphics[width=6cm]{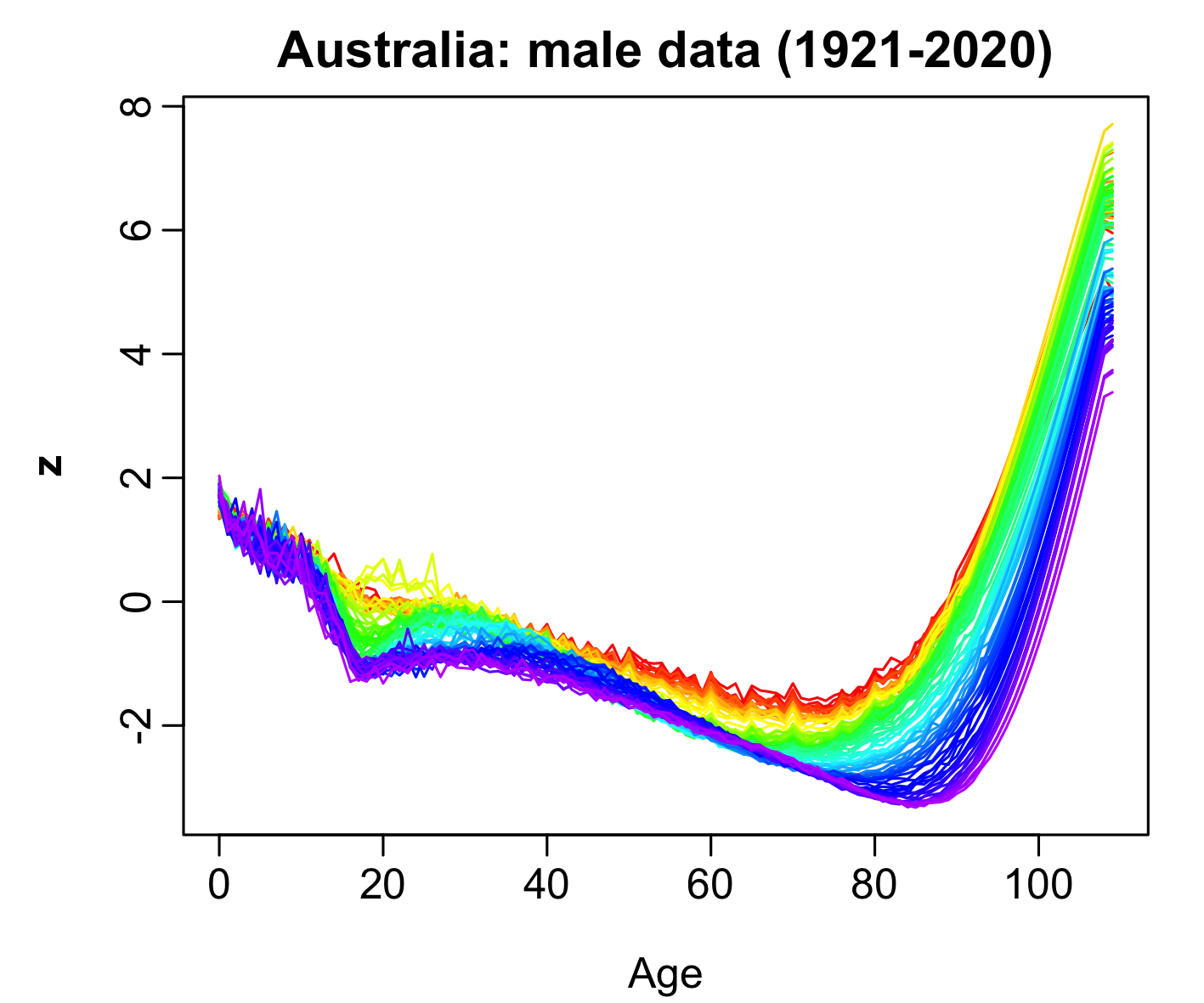}}
\subfloat[1$\textsuperscript{st}$ set of principal component]
{\includegraphics[width=6cm]{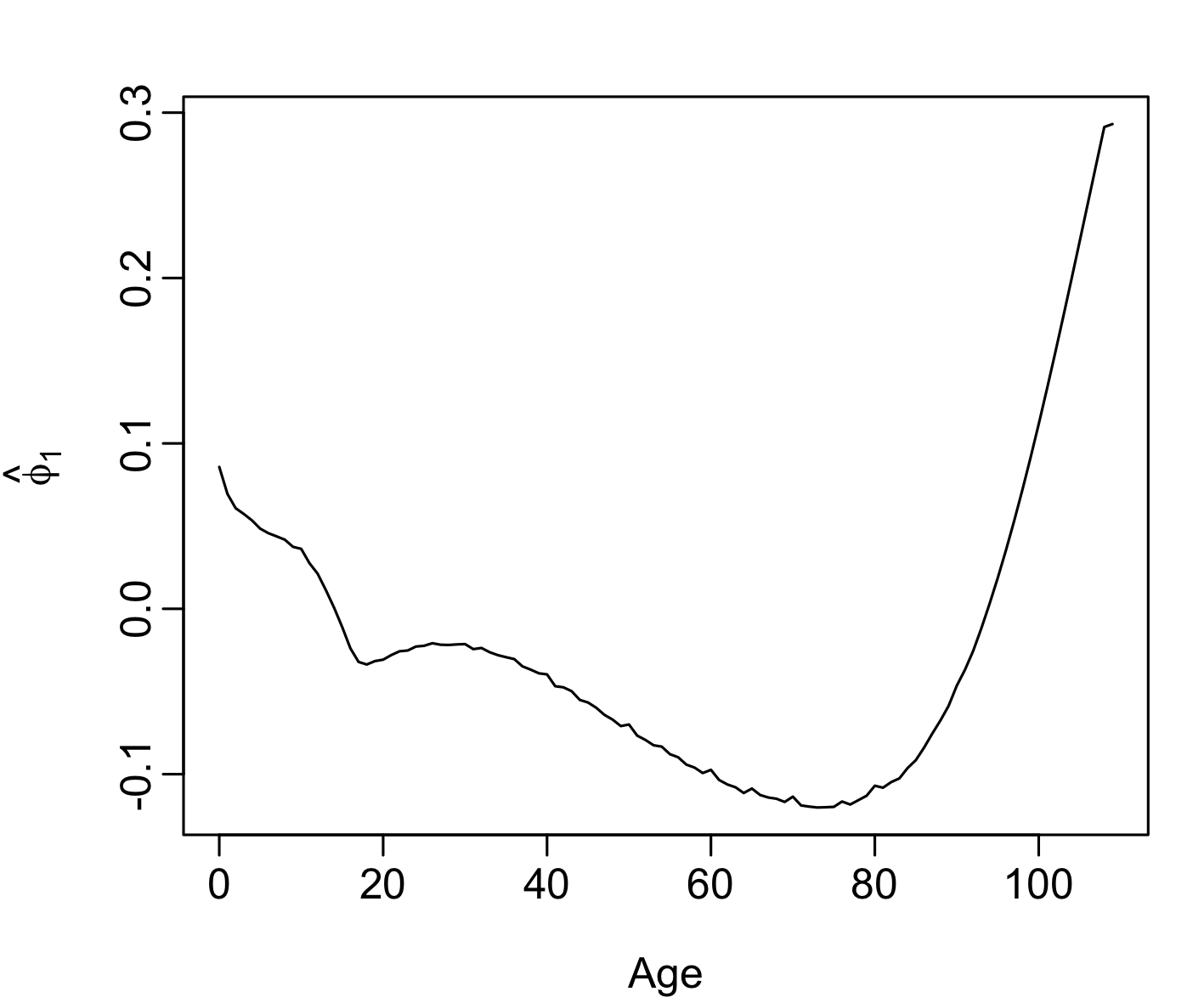}}
\subfloat[2$\textsuperscript{nd}$ set of principal component]
{\includegraphics[width=6cm]{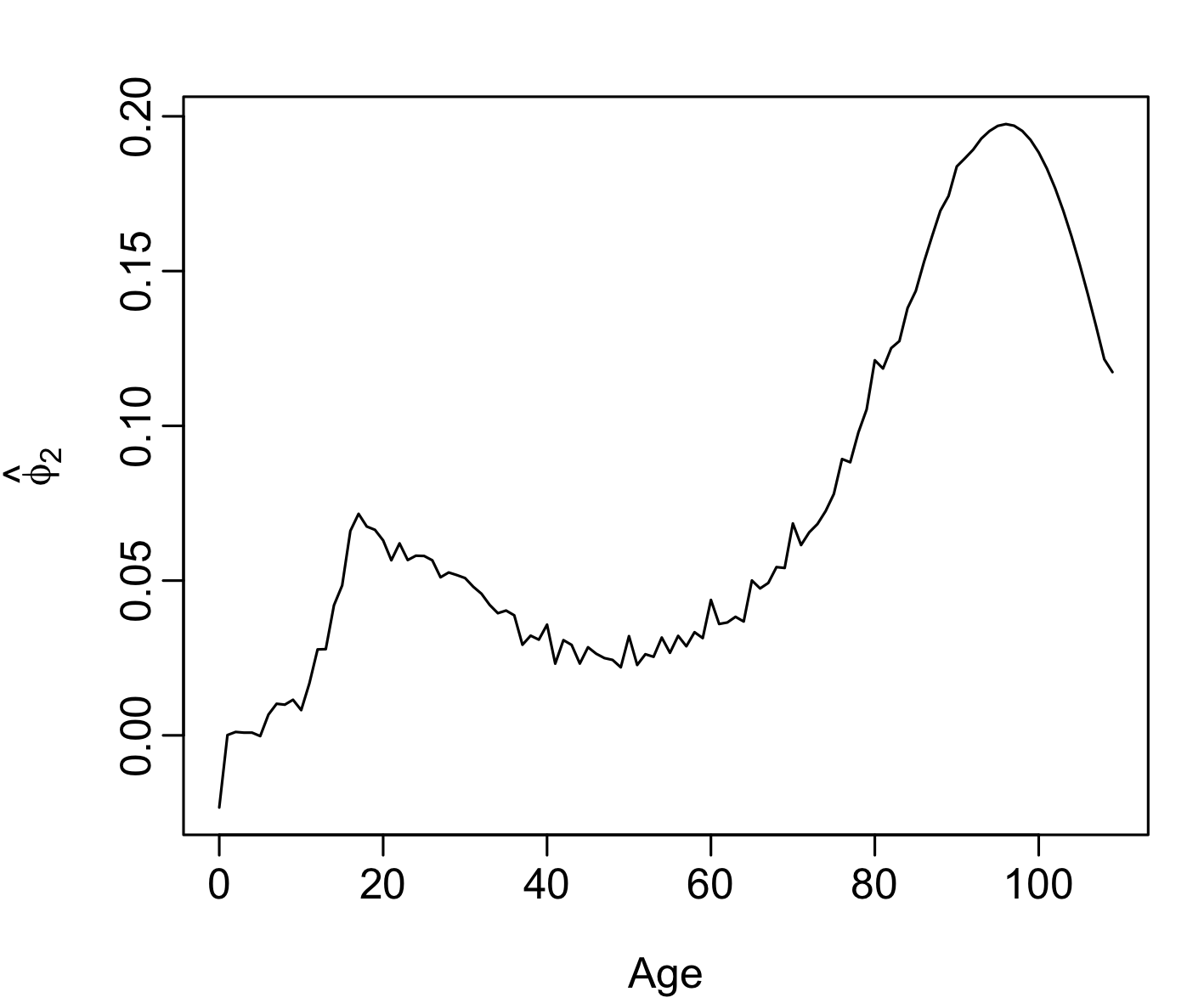}}
\\
\subfloat[Forecast life-table death counts]
{\includegraphics[width=6cm]{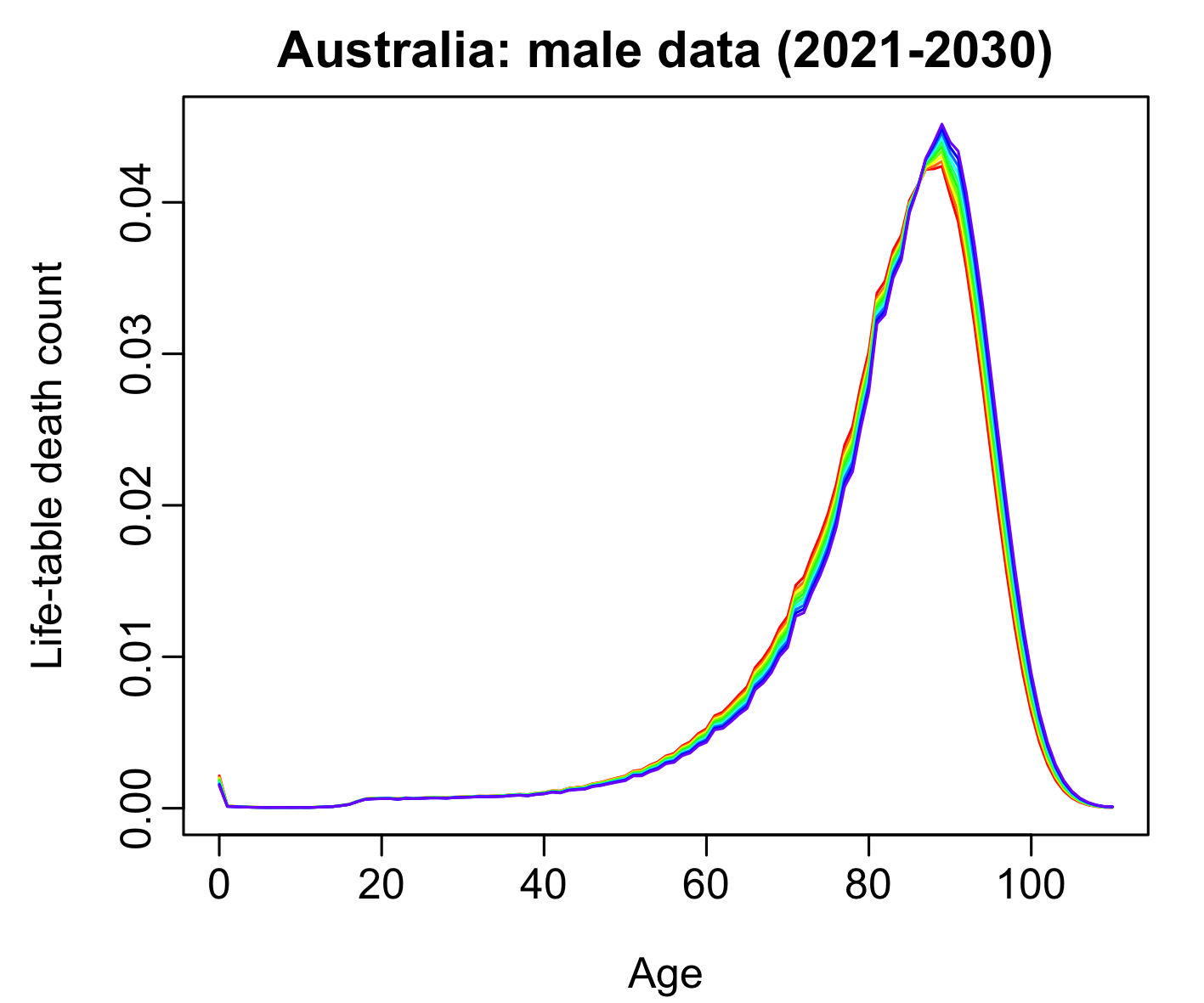}\label{fig:4d}}
\subfloat[Forecast 1$\textsuperscript{st}$ set of scores]
{\includegraphics[width=6cm]{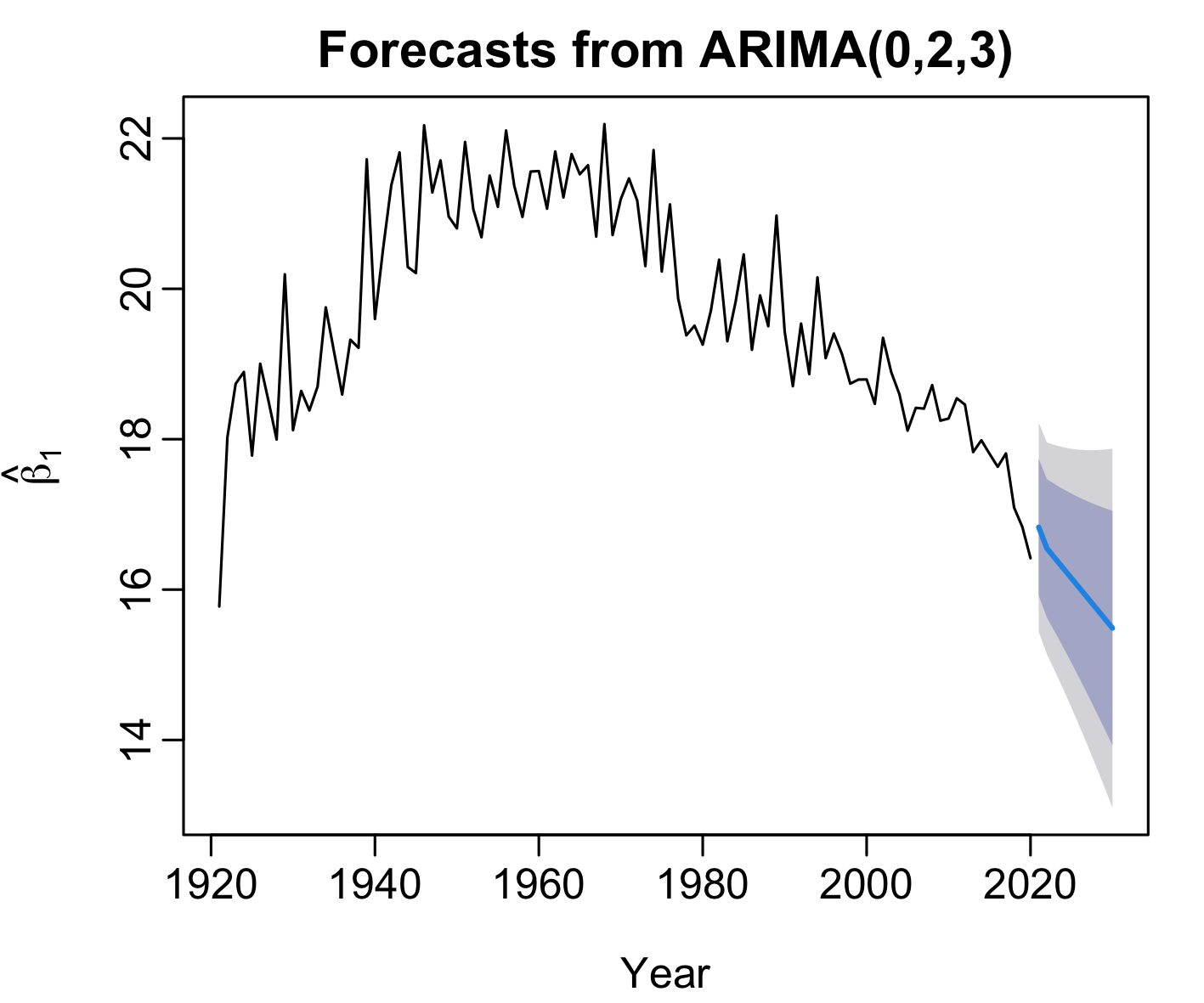}}
\subfloat[Forecast 2$\textsuperscript{nd}$ set of scores]
{\includegraphics[width=6cm]{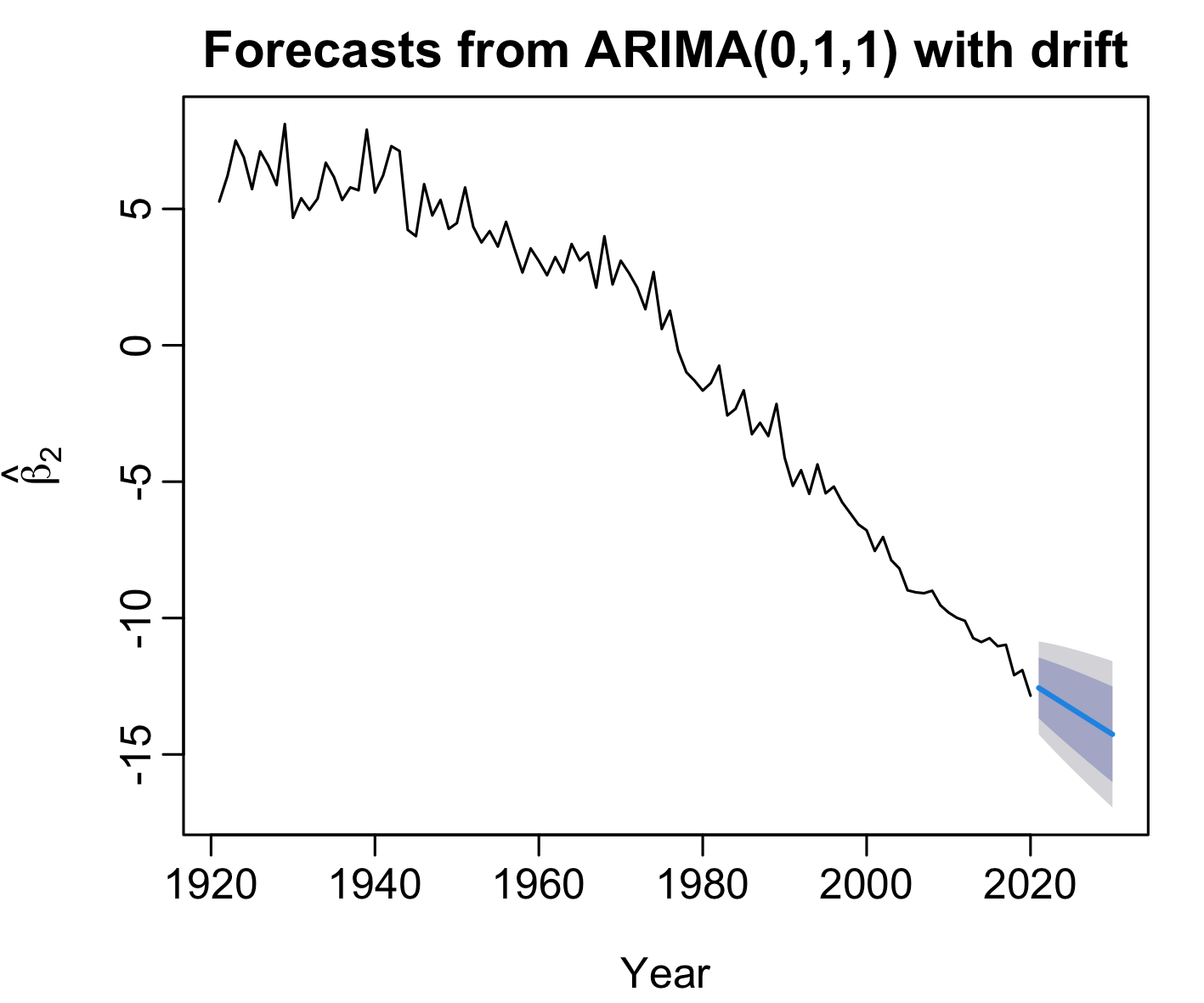}}
\caption{\small Elements of the $\alpha$-transformation for analysing the male age-specific life-table death counts in Australia. We present the first two principal components and their associated principal component scores.}\label{fig:4}
\end{figure}

In Figure~\ref{fig:4}, we display the $\alpha$-transformed data of male life-table death counts, where $\alpha = 0.0528$ on the basis of the Kullback-Leibler divergence with $H=10$. In contrast to the female life-table death counts, the estimated principal components exhibit different shape patterns. Similar to the female life-table death counts, the forecast principal component scores show an increasing trend in the future, indicating a continuation of the recent compression of life-table death counts. By multiplying the forecast principal component scores with the estimated functional principal components and adding the mean function, we produce 10-steps-ahead forecasts of life-table death counts shown in Figure~\ref{fig:4d}. The forecast age distribution of death counts is negatively skewed.

Apart from the graphical display, we can measure the in-sample goodness-of-fit of the $\alpha$ transformation to the observed life-table death counts via an $R^2$ criterion and root mean squared error (RMSE). They can be defined as
\begin{align*}
R^2 &= 1- \frac{\sum^{111}_{x=1}\sum^{100}_{t=1}(d_{t,x}-\widehat{d}_{t,x})^2}{\sum^{111}_{x=1}\sum^{100}_{t=1}(d_{x,t} - \overline{d}_x)^2} \\
\text{RMSE} &= \sqrt{\frac{1}{111\times 100}\sum^{111}_{x=1}\sum^{100}_{t=1}(d_{t,x} - \widehat{d}_{t,x})^2}
\end{align*}
where $d_{t, x}$ denotes the observed age-specific life-table death count for age $x$ in year $t$, and $\widehat{d}_{t,x}$ denotes its estimate. In Table~\ref{tab:1}, we compare the values of $R^2$ and RMSE between the ilr method and $\alpha$ transformation. For both datasets, the eigenvalue ratio criterion selects the number of components $K=2$. The $\alpha$ transformation comparably improves the goodness of fit.
\begin{table}[!htb]
\centering
\tabcolsep 0.37in
\caption{\small The values of $R^2$ and in-sample RMSE between the ilr method and $\alpha$ transformation for the Australian female and male life-table death counts.}\label{tab:1}
\begin{tabular}{@{}llllll@{}}
\toprule
Method & \multicolumn{2}{c}{$R^2$} & & \multicolumn{2}{c}{RMSE} \\
\cmidrule{2-3}\cmidrule{5-6}
	& Female & Male & & Female & Male \\
\midrule
ilr 					& 0.9953 & 0.9911 & & 0.00111 &  0.00141 \\
$\alpha$ transformation 	& 0.9968 & 0.9915 & & 0.00091 & 0.00138 \\
\bottomrule
\end{tabular}
\end{table}

\subsection{Model comparison}

The ilr and eda are special cases of the $\alpha$ transformation. In Section~\ref{sec:4.1}, we also include the clr method without pre-multiplying the Helmert sub-matrix in~\eqref{eq:alpha-transform}. In Section~\ref{sec:4.2}, we compare our methods with the maximum entropy model of \cite{PLC19} based on a set of finite moments.

\subsubsection{Centered log-ratio method}\label{sec:4.1}

There are two versions of the clr:
\begin{inparaenum}
\item[1)] The eigenvalue ratio criterion is a data-driven way of selecting the number of retained components, and it allows the possibility of having more components to improve the in-sample goodness-of-fit. Instead of the random walk with drift, we prefer the ARIMA model as the forecasting method because it can handle nonstationary series.
\item[2)] As advocated in \cite{HBY13}, there is a little price to pay for potentially overfitting by choosing a large number of $K$, such as $K=6$.
\end{inparaenum}

\subsubsection{Maximum entropy mortality (MEM) model}\label{sec:4.2}

There are several attempts to model the age distribution of death counts by parametric mortality laws. \cite{DSS01} used the \citeauthor{HP80}'s \citeyearpar{HP80} model to fit age-specific life-table death counts with Bayesian methods, and \cite{MSZ18} models mortality by fitting a half-normal and a skew-bimodal-normal distribution to the observed age-at-death density function. While these approaches are parametric, \cite{PLC19} consider a nonparametric alternative based on statistical moments. The forecast life-table death counts for a population can be determined by extrapolating finite statistical moments derived from the observed data. 

The extrapolation is achieved through multivariate time series models, such as multivariate random walk with drift. The age distribution of death counts is then estimated in time from the predicted moments. In practice, the first four moments are sufficient to characterise a distribution.

\subsection{Point forecast error measures}\label{sec:4.3}

Since life-table death counts can be considered a probability density function, we study some density evaluation measures. These measures include the discrete version of the Kullback-Leibler divergence \citep{KL51} and the square root of the Jensen-Shannon divergence \citep{Shannon48}. For two probability density functions, denoted by $\bm{d}_{m+\xi}$ and $\widehat{\bm{d}}_{m+\xi}$, the \textit{symmetric} version of the Kullback-Leibler divergence is defined as
\begin{align*}
\text{KLD}(h) =& \ D_{\text{KL}}\big(\bm{d}_{m+\xi}||\widehat{\bm{d}}_{m+\xi}\big) + D_{\text{KL}}\big(\widehat{\bm{d}}_{m+\xi}||\bm{d}_{m+\xi}\big) \\
		     =& \ \frac{1}{111\times (H+1-h)}\sum^{H}_{\xi=h}\sum^{111}_{x=1}d_{m+\xi, x}\cdot \big(\ln d_{m+\xi, x}-\ln \widehat{d}_{m+\xi,x}\big) \\
		       &+ \frac{1}{111\times (H+1-h)}\sum^{H}_{\xi=h}\sum^{111}_{x=1}\widehat{d}_{m+\xi, x}\cdot \big(\ln \widehat{d}_{m+\xi, x}-\ln d_{m+\xi,x}\big),
\end{align*}
which is non-negative, for $h=1,2,\dots,H$. Let $m$ denote the number of data samples in the training set for selecting the optimal $\alpha$ in the validation set. With a slight abuse of notation, $m$ can also represent the data in the training sample when examining the out-of-sample forecast accuracy.

An alternative metric is the Jensen-Shannon divergence, defined by
\[
\text{JSD}(h) = \frac{1}{2}D_{\text{KL}}\Big(\bm{d}_{m+\xi}||\bm{\delta}_{m+\xi}\Big) + \frac{1}{2}D_{\text{KL}}\Big(\widehat{\bm{d}}_{m+\xi}||\bm{\delta}_{m+\xi}\Big),
\]
where $\bm{\delta}_{m+\xi}$ measures a common quantity between $\bm{d}_{m+\xi}$ and $\widehat{\bm{d}}_{m+\xi}$. As in \cite{SH20}, we consider two cases: the arithmetic mean and the geometric mean
\begin{align*}
\bm{\delta}_{m+\xi} &= \frac{1}{2}\big(\bm{d}_{m+\xi}+\widehat{\bm{d}}_{m+\xi}\big), \\
\bm{\delta}_{m+\xi} &= \sqrt{\bm{d}_{m+\xi}\widehat{\bm{d}}_{m+\xi}}.
\end{align*}
Let $\text{JSD}^{\text{a}}(h)$ be the Jensen-Shannon divergence with the arithmetic mean, while $\text{JSD}^{\text{g}}(h)$ be the Jensen-Shannon divergence with the geometric mean.

For $h=1,2,\dots,H$, we compute the averaged KLD, JSD$^{a}$ and JSD$^{g}$ defined as
\begin{align*}
\text{KLD} &= \frac{1}{H}\sum^{H}_{h=1}\text{KLD}(h), \\
\text{JSD}^a &= \frac{1}{H}\sum^{H}_{h=1}\text{JSD}^a(h), \\
\text{JSD}^g &= \frac{1}{H}\sum^{H}_{h=1}\text{JSD}^g(h).
\end{align*}

\subsection{Point forecast results}\label{sec:4.4}

As explained in \cite{Booth06}, there exist three approaches to forecasting demographic processes, namely extrapolation, expectation and theory-based structural modelling involving exogenous variables. As our method belongs to a time-series extrapolation approach, we focus on a short-term forecast horizon, such as $H=10$ or 20. We implement an \textit{expanding window} approach to evaluate and compare point and interval forecasts. The expanding window is a forecasting technique where we iteratively increase the size of the training sample to make our predictions. 

Based on the KLD, JSD$^a$ and JSD$^g$, we evaluate and compare the point forecast accuracy of the $\alpha$ transformation and its two special cases (ilr and eda). We consider forecasting each set of the estimated principal component scores by the selected ARIMA model based on the correct Akaike information criterion for each horizon $h=1,2,\dots,H$.

\begin{table}[!htb]
\centering
\tabcolsep 0.083in
\caption{\small For two different forecast horizons $H=10$ and 20, we present a comparison of the point forecast accuracy, measured by the KLD, JSD$^a$ and JSD$^g$, among the $\alpha$ transformation, isometric log-ratio (ilr), and Euclidean data analysis (eda), using the holdout sample of the Australian female and male life-table death counts. We also include the clr method and the MEM of \cite{PLC19} for comparison. The smallest errors among the methods are highlighted in bold for $H=10$ and 20.}\label{tab:2}
\begin{tabular}{@{}llllllllllll@{}}
\toprule
	&	&	& \multicolumn{4}{c}{Eigenvalue ratio} &  \multicolumn{4}{c}{$K=6$} & \\
Sex & Criterion & $H$ & $\alpha$ & ilr & eda & clr & $\alpha$ & ilr & eda & clr  & MEM \\  \midrule
F 	& KLD 	& 10   	& \textBF{0.0062} 	& 0.0099 	& 0.0925 	& 0.0109  & \textBF{0.0036} 	& \textBF{0.0036} 	& 0.0104  & 0.0037 & 0.0053 \\ 
	&		& 20 		& \textBF{0.0101}  	&  0.0134  & 0.0729  & 0.0150 	& \textBF{0.0045}  	&  0.0056   		&  0.0241 & 0.0059 & 0.0123 \\
\\
& $\text{JSD}^{\text{a}}$ 	& 10 & \textBF{0.0016} & 0.0025 & 0.0127 & 0.0027 & \textBF{0.0009} & \textBF{0.0009} & 0.0023  & 0.0009 & 0.0013 \\ 
& 					& 20 & \textBF{0.0025} & 0.0033 & 0.0114 & 0.0037 & \textBF{0.0014} & \textBF{0.0014} & 0.0055 & 0.0015 & 0.0030 \\
\\
& $\text{JSD}^{\text{g}}$ 	& 10 & \textBF{0.0016} & 0.0025 & 0.0131 & 0.0027 & \textBF{0.0009} & \textBF{0.0009} & 0.0024 & \textBF{0.0009} & 0.0013 \\ 
& 					& 20 & \textBF{0.0025} & 0.0033 & 0.0120 & 0.0037 & \textBF{0.0014} & \textBF{0.0014} & 0.0060 & 0.0015 		  & 0.0031 \\
\midrule
M 	& KLD 	& 10 & 0.0174 	& 0.0204 & 0.0479  & 0.0197 & 0.0077 	& 0.0091 	& 0.0298  & 0.0082 & \textBF{0.0046} \\
	& 		& 20 	& 0.0418  & 0.0453  &  0.3564 & 0.0431 & 0.0192  	&  0.0213  &  0.3125 & 0.0292 & \textBF{0.0084} \\
\\
& $\text{JSD}^{\text{a}}$ 	& 10   	& 0.0042 & 0.0050 & 0.0104  & 0.0049 & 0.0020 & 0.0022 & 0.0067  & 0.0020 & \textBF{0.0011} \\ 
	& 				& 20 		& 0.0103 & 0.0112 & 0.0235 & 0.0106 & 0.0048  & 0.0052 & 0.0201 & 0.0072 & \textBF{0.0021} \\
\\
& $\text{JSD}^{\text{g}}$ 	& 10   	& 0.0042 & 0.0051 & 0.0124 & 0.0049 & 0.0020 & 0.0023 & 0.0076 & 0.0020 & \textBF{0.0011} \\ 
& 					& 20 		& 0.0104 & 0.0113 & 0.0297 & 0.0107 & 0.0049 & 0.0053 & 0.0249 & 0.0073 & \textBF{0.0021} \\
\bottomrule
\end{tabular}
\end{table}

The detailed results are presented in Table~\ref{tab:2} for the five methods, where we average over the ten forecast horizons. As the forecast horizon increases, the forecast errors generally become larger. This phenomenon reflects the increasing uncertainties associated with the model and estimated parameters. The $\alpha$-transformation method performs the best among the five methods for the female life-table death counts. In contrast, the MEM produces the most accurate forecasts for the male life-table death counts. The $\alpha$-transformation and ilr methods tend to produce more accurate forecasts than the eda method as it respects the non-negativity and summability constraints. The $\alpha$-transformation and ilr methods can also adapt to temporal changes in the age distribution of the life-table death counts over the years. Between the ilr and clr, they perform similarly in terms of point forecast accuracy. Between the two methods of selecting $K$, having a larger number of components not only improves model fitting but also improves point forecast accuracy.

\commHS{For comparison, we also consider applying the Lee-Carter model to forecast age-specific mortality rates directly and then transforming the forecasts into the forecasted life-table death counts. Computationally, the conversion between rate and life-table death counts can be achieved via the \textit{LifeTable} function in the MortalityLaws package \citep{Pascariu24}. Since our objective is to model life-table death counts at ages $0, 1,\dots, 109, 110+$, it is necessary to smooth the original data via penalized regression spline with monotonic constraint. Computationally, the smoothing step can be achieved via the \textit{smooth.demogdata} function in demography package \citep{Hyndman23}. Although this comparison does not impact our recommendation, it presents an interesting research direction to explore various mortality instruments.}
\begin{table}[!htb]
\centering
\tabcolsep 0.3in
\caption{\small{For comparison, we also consider forecasting the age-specific mortality rates via the Lee-Carter model. Computationally, this can achieved via the \textit{lca} function in the demography package. From the forecast mortality rates, we turn them into the forecasted life-table death counts.}}
\begin{tabular}{@{}llll@{}}
\toprule
& & \multicolumn{2}{c}{Sex}  \\
Criterion & $H$ & F & M \\
\midrule
KLD & 10 & 0.0050 & 0.0114 \\
 	& 20 & 0.0077 & 0.0253 \\
\\	
$JSD^{\text{a}}$  & 10 &  0.0012 & 0.0028 \\
	& 20 & 0.0019 & 0.0063 \\
\\
$JSD^{\text{g}}$  & 10 & 0.0012 & 0.0028 \\
 	& 20 & 0.0019 & 0.0063 \\
\bottomrule
\end{tabular}
\end{table}

In Table~\ref{tab:3}, we present the point forecast results using a \textit{rolling window} scheme. With the rolling-window scheme, the training sample considers the more recent data and the size of the training sample remains the same. For the female series, the $\alpha$-transformation, ilr and clr perform similarly, outperforming the forecasts obtained from the MEM. For the male series, the MEM provides the best point forecast accuracy. These findings are aligned with our previous findings using the expanding window scheme.

\begin{center}
\tabcolsep 0.075in
\begin{longtable}{@{}llllllllllll@{}}
\caption{\small For two different forecast horizons $H=10$ and 20, we present a comparison of the point forecast accuracy using the rolling-window scheme, measured by the KLD, JSD$^a$ and JSD$^g$, among the $\alpha$ transformation, isometric log-ratio (ilr), and Euclidean data analysis (eda), using the holdout sample of the Australian female and male life-table death counts. We also include the clr method and the MEM of \cite{PLC19} for comparison. The smallest errors among the methods are highlighted in bold for $H=10$ and 20.}\label{tab:3} \\
\toprule
	&	& \multicolumn{5}{c}{$K=6$ (Female)} &  \multicolumn{5}{c}{$K=6$ (Male)} \\
Criterion & $H$ & $\alpha$ & ilr & eda & clr & MEM & $\alpha$ & ilr & eda & clr  & MEM \\  \midrule
\endfirsthead
&	& \multicolumn{5}{c}{$K=6$ (Female)} &  \multicolumn{5}{c}{$K=6$ (Male)} \\
Criterion & $H$ & $\alpha$ & ilr & eda & clr & MEM & $\alpha$ & ilr & eda & clr  & MEM \\  \midrule
\endhead
\hline \multicolumn{12}{r}{{Continued on next page}} \\
\endfoot
\endlastfoot
 KLD 				& 10 & 0.0033 & 0.0033 & 0.0094 & 0.0033 & 0.0048 & 0.0079 & 0.0094 & 0.0287 & 0.0094 & \textBF{0.0043} \\
					& 20 	& 0.0050 & 0.0050 & 0.0227 & 0.0050 & 0.0102 & 0.0203 & 0.0208 & 0.2694 & 0.0208 & \textBF{0.0075} \\
\\
 $\text{JSD}^{\text{a}}$ 	& 10 & 0.0008 & 0.0008 & 0.0022 & 0.0008 & 0.0012 & 0.0020 & 0.0023 & 0.0064 & 0.0023 & \textBF{0.0010} \\
 					& 20 &  0.0012 & 0.0012 & 0.0052 & 0.0012 & 0.0025 & 0.0049 & 0.0051 & 0.0184 & 0.0051 & \textBF{0.0018} \\
\\
 $\text{JSD}^{\text{g}}$ 	& 10 &  0.0008 & 0.0008 & 0.0023 & 0.0008 & 0.0012 & 0.0020 & 0.0023 & 0.0072 & 0.0023 & \textBF{0.0011} \\
 					& 20 &  0.0012 & 0.0012 & 0.0056 & 0.0012 & 0.0025 & 0.0050 & 0.0052 & 0.0229 & 0.0052 & \textBF{0.0019} \\
\bottomrule
\end{longtable}
\end{center}

\vspace{-.6in}

\subsection{Interval forecast error measures}\label{sec:4.5}

We consider the absolute difference between empirical and nominal coverage probabilities and the mean interval score of \cite{GR07} to evaluate the interval forecast accuracy. For each year in the forecasting period, the $h$-step-ahead prediction intervals are computed at the $100(1-\gamma)\%$ nominal coverage probability. For constructing $100(1-\gamma)\%$ prediction intervals, let us denote $\widehat{d}_{m+\xi,x}^{\text{lb}}$ and $\widehat{d}_{m+\xi,x}^{\text{ub}}$ as the lower and upper bounds, respectively.  We compute the empirical coverage probability (ECP), defined as
\[
\text{ECP}_{\gamma}(h) = 1 - \frac{1}{111\times (H+1-h)}\sum^{H}_{\xi=h}\sum^{111}_{x=1}\left[\mathds{1}\{d_{m+\xi,x}<\widehat{d}_{m+\xi,x}^{\text{lb}}\} + \mathds{1}\{d_{m+\xi,x}>\widehat{d}_{m+\xi,x}^{\text{ub}}\}\right].
\]
While the ECP reveals over- or under-estimation of the nominal coverage probability, it is not an accuracy criterion due to the possible cancellation. As an alternative, the CPD is defined as
\[
\text{CPD}_{\gamma}(h) = \left|\text{ECP}_{\gamma}(h) - (1-\gamma)\right|.
\]
The smaller the value of CPD is, the more accurate interval forecast accuracy the method produces. Despite the ECP and CPD being measures of interval forecast accuracy, neither consider the sharpness of the prediction intervals, i.e., the distance between the lower and upper bounds. To address this problem, \cite{GR07} introduce a scoring rule for the interval forecasts at time point $d_{n+\xi,x}$:
\begin{align*}
S_{\gamma,\xi}\big(\widehat{d}_{m+\xi,x}^{\text{lb}}, \widehat{d}_{m+\xi,x}^{\text{ub}}, d_{m+\xi, x}\big)= \big(\widehat{d}_{m+\xi,x}^{\text{ub}}-\widehat{d}_{m+\xi,x}^{\text{lb}}\big) &+ \frac{2}{\gamma}\big(\widehat{d}_{m+\xi,x}^{\text{lb}}-d_{m+\xi,x}\big)\mathds{1}\{d_{m+\xi,x}<\widehat{d}_{m+\xi,x}^{\text{lb}}\} \\
                  &+ \frac{2}{\gamma}\big(d_{m+\xi,x}-\widehat{d}_{m+\xi,x}^{\text{ub}}\big)\mathds{1}\{d_{m+\xi,x}>\widehat{d}_{m+\xi,x}^{\text{ub}}\}.
\end{align*}
The interval score rewards a narrow prediction interval if and only if the true observation lies within the prediction interval. The optimal interval score is achieved when $d_{m+\xi,x}$ lies between $\widehat{d}_{m+\xi,x}^{\text{lb}}$ and $\widehat{d}_{m+\xi,x}^{\text{ub}}$ and the distance between $\widehat{d}_{m+\xi,x}^{\text{lb}}$ and $\widehat{d}_{m+\xi,x}^{\text{ub}}$ is minimal for age $x$.

For different ages and years in the forecasting period, the mean interval score is defined by
\[
S_{\gamma}(h) = \frac{1}{111\times (H+1-h)}\sum^{H}_{\xi=h}\sum^{111}_{x=1}S_{\gamma,\xi}\Big(\widehat{d}_{m+\xi, x}^{\text{lb}}, \widehat{d}_{m+\xi, x}^{\text{ub}}; d_{m+\xi, x}\Big),
\]
where $S_{\gamma,\xi}\Big(\widehat{d}_{m+\xi, x}^{\text{lb}}, \widehat{d}_{m+\xi, x}^{\text{ub}}; d_{m+\xi, x}\Big)$ denotes the interval score at the $\xi$\textsuperscript{th} curve in the validation or testing set. 

For $h=1,2,\dots, H$, we compute the averaged CPD and interval scores:
\begin{equation*}
\text{CPD}_{\gamma} = \frac{1}{H}\sum^H_{h=1}\text{CPD}_{\gamma}(h), \qquad
S_{\gamma} = \frac{1}{H}\sum^H_{h=1}S_{\gamma}(h).
\end{equation*}

\subsection{Interval forecast results}\label{sec:4.6}

Using the \textit{expanding window} approach, we also evaluate and compare the interval forecast accuracy of the $\alpha$ transformation, ilr and eda, based on coverage probability difference (CPD) and $S_{\gamma}$. The detailed results are presented in Table~\ref{tab:4} for the three methods. Between the $\alpha$ transformation and ilr, the former is more flexible and can provide improved forecast accuracy subject to a finer tuning of the $\alpha$ parameter. For the female data, the $\alpha$ transformation generally produces the smallest CPD values and mean interval scores for $\alpha=0.05$ and 0.2. For the male data, the ilr method generally produces the smallest CPD values and mean interval scores. This result further highlights the importance of selecting an optimal $\alpha$ value and its difficulty associated with the relatively volatile male data. Because the eda ignores the compositional constraints, it shows the worst interval forecast accuracy. When the number of components is $K=6$, it improves the interval forecast accuracy, especially for the eda method. We also computed the pointwise prediction intervals obtained from the MEM. Because each of the lower and upper bounds sums up to the radix, this constraint can result in cross-over. Hence, it produces inferior interval forecast accuracy. \commHS{Because of this cross-over effect, the conversion from mortality rate forecasts to forecasted life-table death count will also have inferior forecast accuracy.}

\begin{center}
\tabcolsep 0.15in
\begin{longtable}{@{}llllllllll@{}}
\caption{\small Comparison of the interval forecast accuracy, as measured by the CPD$_{\gamma}$ and $S_{\gamma}$, among the $\alpha$ transformation, isometric log-ratio (ilr) and Euclidean data analysis (eda), using the holdout sample of the Australian female and male life-table death counts.}\label{tab:4} \\
\toprule
	  	&	& & \multicolumn{3}{c}{Eigenvalue ratio}   			& \multicolumn{3}{c}{$K=6$} &  \\
Sex 		& Criterion 		& $H$ 	& alpha 	& ilr 	 & eda  	& alpha 	& ilr 		& eda & MEM \\\midrule
\endfirsthead
\toprule
	  	&	& & \multicolumn{3}{c}{Eigenvalue ratio}   			& \multicolumn{3}{c}{$K=6$} &  \\
Sex 		& Criterion 		& $H$ 	& alpha 	& ilr 	 & eda  	& alpha 	& ilr 		& eda & MEM \\\midrule
\endhead
\hline \multicolumn{10}{r}{{Continued on next page}} \\
\endfoot
\endlastfoot
F 		&  CPD$_{0.2}$  	&  10 	& 0.0486 & 0.0765 & 0.2652 & 0.0497 & 0.1065 & 0.0598 & 0.8171 \\ 
		&				&  20		& 0.0744 & 0.0969 & 0.1490 & 0.0553 & 0.1007 & 0.0857 & 0.8199  \\ 
\\
		& $S_{0.2}$  		&  10 	& 0.0029 & 0.0034 & 0.0074 & 0.0022 & 0.0027 & 0.0034 & 0.0100 \\ 
		&				& 20 		& 0.0033 & 0.0037 & 0.0069 & 0.0024 & 0.0032 & 0.0045 & 0.0149 \\ 
\\
		&  CPD$_{0.05}$  	&  10 	& 0.0218 & 0.0341 & 0.1165 & 0.0367 & 0.0367 & 0.0456 & 0.8702  \\  
		&				&  20		& 0.0334 & 0.0394 & 0.0761 & 0.0308 & 0.0435 & 0.0433 & 0.8513 \\ 
\\
		&  $S_{0.05}$ 		&  10 	& 0.0034 & 0.0046 & 0.0130 & 0.0036 & 0.0038 & 0.0060  & 0.0543 \\
		&				&  20		& 0.0042 & 0.0054 & 0.0113 & 0.0034 & 0.0047 & 0.0065  & 0.0828 \\ 
\midrule
M 		&  CPD$_{0.2}$ 	&  10 	& 0.1311 & 0.1099 & 0.2562 & 0.1028 & 0.0854 & 0.1187  & 0.8351 \\
		&				&  20		& 0.3417 & 0.1390 & 0.4503 & 0.1511 & 0.0620 & 0.2653 & 0.8443 \\ 
\\	
		& $S_{0.2}$ 		&  10 	& 0.0053 & 0.0052 & 0.0068 & 0.0032 & 0.0030 & 0.0048  & 0.0096 \\
		&				&  20		& 0.0106 & 0.0089 & 0.0119 & 0.0052 & 0.0040 & 0.0090 & 0.0144 \\ 
\\
		& CPD$_{0.05}$ 	&  10 	& 0.0841 & 0.0421 & 0.2030 & 0.0418 & 0.0397 & 0.0612 & 0.9145 \\
		&				&  20		& 0.2008 & 0.1168 & 0.3350 & 0.0956 & 0.0353 & 0.1533 & 0.9239 \\ 
\\
		& $S_{0.05}$  		& 10 		& 0.0066 & 0.0060 & 0.0128 & 0.0040 & 0.0039 & 0.0079  & 0.0512 \\
		&				& 20		& 0.0142 & 0.0136 & 0.0238 & 0.0086 & 0.0051 & 0.0167 & 0.0794 \\ 
\bottomrule
\end{longtable}
\end{center}

\vspace{-.2in}

As a sensitivity analysis, we conduct the interval forecast accuracy comparison using the rolling window approach and the results are shown in Table~\ref{tab:5}. For modelling the female series, the alpha transformation produces the smallest interval forecast errors, measured by the CPD and interval score. However, for the male series, the ilr method is advantageous and is our recommended approach. A partial reason is that the selected $\alpha$ parameter from the validation data set is sub-optimal for the testing data set. Between the eigenvalue ratio and $K=6$, the latter one produces comparably smaller interval forecast errors.

\begin{center}
\tabcolsep 0.155in
\begin{longtable}{@{}llllllllll@{}}
\caption{\small Comparison of the interval forecast accuracy, as measured by the CPD$_{\gamma}$ and $S_{\gamma}$, among the $\alpha$ transformation, isometric log-ratio (ilr) and Euclidean data analysis (eda), using the rolling window scheme.}\label{tab:5} \\
\toprule
	  	&	& & \multicolumn{3}{c}{Eigenvalue ratio}   			& \multicolumn{3}{c}{$K=6$} &  \\
Sex 		& Criterion 		& $H$ 	& alpha 	& ilr 	 & eda  	& alpha 	& ilr 		& eda & MEM \\\midrule
\endfirsthead
\toprule
	  	&	& & \multicolumn{3}{c}{Eigenvalue ratio}   			& \multicolumn{3}{c}{$K=6$} &  \\
Sex 		& Criterion 		& $H$ 	& alpha 	& ilr 	 & eda  	& alpha 	& ilr 		& eda & MEM \\\midrule
\endhead
\hline \multicolumn{10}{r}{{Continued on next page}} \\
\endfoot
\endlastfoot
F 		&  CPD$_{0.2}$  	&  10 	& 0.0627 & 0.0785 & 0.2797 & 0.0453 & 0.0976 & 0.0667 & 0.8179 \\ 
		&				&  20		& 0.0910 & 0.0958 & 0.1752 & 0.0660 & 0.0757 & 0.0912 & 0.8277 \\ 
\\
		& $S_{0.2}$  		&  10 	& 0.0029 & 0.0033 & 0.0073 & 0.0023 & 0.0026 & 0.0034 & 0.0092 \\ 
		&				& 20 		& 0.0033 & 0.0036 & 0.0074 & 0.0025 & 0.0033 & 0.0045 & 0.0132 \\ 
\\
		&  CPD$_{0.05}$  	&  10 	& 0.0222 & 0.0319 & 0.1254 & 0.0367 & 0.0358 & 0.0450 & 0.8798 \\ 
		&				&  20		& 0.0325 & 0.0378 & 0.0870 & 0.0328 & 0.0396 & 0.0445 & 0.8833 \\ 
\\
		&  $S_{0.05}$ 		&  10 	& 0.0037 & 0.0046 & 0.0127 & 0.0036 & 0.0038 & 0.0057 & 0.0498 \\  
		&				&  20		& 0.0041 & 0.0052 & 0.0128 & 0.0035 & 0.0047 & 0.0067 & 0.0723 \\ 
\midrule
M 		&  CPD$_{0.2}$ 	&  10 	& 0.1655 & 0.1387 & 0.2441 & 0.1143 & 0.0954 & 0.0981 & 0.8307 \\
		&				&  20		& 0.3532 & 0.1642 & 0.4284 & 0.1600 & 0.0926 & 0.2144 & 0.8387  \\ 
\\	
		& $S_{0.2}$ 		&  10 	& 0.0053 & 0.0053 & 0.0069 & 0.0033 & 0.0031 & 0.0048 & 0.0090 \\
		&				&  20		& 0.0093 & 0.0093 & 0.0151 & 0.0054 & 0.0043 & 0.0089 & 0.0130 \\ 
\\
		& CPD$_{0.05}$ 	&  10 	& 0.0718 & 0.0580 & 0.2155 & 0.0687 & 0.0446 & 0.0573 & 0.9152 \\ 
		&				&  20		& 0.2612 & 0.1302 & 0.3579 & 0.1162 & 0.0387 & 0.1388 & 0.9214 \\ 
\\
		& $S_{0.05}$  		& 10 		& 0.0067 & 0.0064 & 0.0133 & 0.0039 & 0.0039 & 0.0080 & 0.0480 \\ 
		&				& 20		& 0.0156 & 0.0155 & 0.0313 & 0.0089 & 0.0053 & 0.0167 & 0.0715 \\ 
\bottomrule
\end{longtable}
\end{center}

\vspace{-.5in}

\section{Concluding comments}\label{sec:5}

We consider the $\alpha$ transformation to model and forecast age-specific life-table death counts in Australia. We evaluate the point and interval forecast accuracies among the $\alpha$ transformation, isometric log-ratio analysis, and Euclidean data analysis for forecasting the age distribution of death counts. Based on the Kullback-Leibler and Jensen-Shannon divergences, the $\alpha$ transformation is recommended for modelling the female data, while the MEM method is advocated for modelling the male data. Subject to an optimal selection of $\alpha$ parameter, the $\alpha$ transformation is more flexible and can provide improved point forecast accuracy than the log-ratio analysis. Similar to the Box-Cox transformation, the $\alpha$ transformation is a natural way of handling zeros in compositional data analysis. In terms of interval forecast accuracy, the $\alpha$ transformation also provides the smallest interval forecast errors, measured by CPD and mean interval score for the female data. For the more volatile male data, the ilr method is advocated. For reproducibility, all the code is available on GitHub: \url{https://github.com/hanshang/alpha_transformation}.

There are a few ways in which this paper could be extended, and we briefly discuss five:
\begin{asparaenum}
\item[1)] In the demographic literature, there are other models for forecasting the age distribution of death, such as the segmented transformation age-at-death distribution from \cite{BC19}.
\item[2)] A robust $\alpha$ transformation may be proposed in the presence of outlying years.
\item[3)] One can extend $\alpha$ transformation to jointly model and forecast the age distribution of death counts for multiple populations.
\item[4)] We considered period life table, which is a useful measure of mortality rates experienced over a given period. As an alternative, one can consider cohort life-table death counts. A cohort life table displays the probability of a person from a given cohort dying at each age over the course of their lifetime. In this setting, a cohort refers to a group of individuals with the same year of birth.
\item[5)] One could consider a probabilistic framework that accounts for compositional structure of the data, such as the Dirichlet distribution \citep[see, e.g.,][]{TS18, TWY+22, GN23}. The use of the Dirichlet distribution is to avoid any transformation and directly model life-table death counts that are easier to interpret.
\end{asparaenum}

\section*{Acknowledgements}

The authors would like to acknowledge the insightful comments from two reviewers that led to a much-improved paper, and fruitful discussions with Professor Denis Allard, who introduced us to $\alpha$ transformation at the KAUST2022 workshop on statistics. The authors thank helpful comments received from the participants at the~2023 SOA Living to 100 symposium, the Population Association of New Zealand conference held at the University of Auckland, and a seminar at the Research School of Finance, Actuarial Studies and Statistics at the Australian National University. The first author acknowledges the fundings from the Australian Research Council under Discovery Project Grant DP230102250 and Future Fellow FT240100338.

\section*{Declarations}

\textbf{Disclosure statement:} The authors report there are no competing interests to declare. \\

\noindent \textbf{Data availability statement:} The datasets are available from \url{https://github.com/hanshang/alpha_transformation}.

\bibliographystyle{apalike}
\bibliography{alpha_transformation.bib}

\begin{thebibliography}{}

\bibitem[Aburto et~al., 2022]{ABB+22}
Aburto, J.~M., Basellini, U., Baudisch, A., and Villavicencio, F. (2022).
\newblock {Drewnowski's index to measure lifespan variation: Revisiting the
  Gini coefficient of the life table}.
\newblock {\em Theoretical Population Biology}, 148:1--10.

\bibitem[Aitchison, 1982]{Aitchison1982}
Aitchison, J. (1982).
\newblock {The statistical analysis of compositional data}.
\newblock {\em Journal of the Royal Statistical Society: Series B},
  44(2):139--177.

\bibitem[Aitchison, 1986]{Aitchison1986}
Aitchison, J. (1986).
\newblock {\em {The Statistical Analysis of Compositional Data}}.
\newblock Chapman \& Hall, London.

\bibitem[Aitchison and Shen, 1980]{AS80}
Aitchison, J. and Shen, S.~M. (1980).
\newblock {Logistic-normal distributions: Some properties and uses}.
\newblock {\em Biometrika}, 67(2):261--272.

\bibitem[Basellini and Camarda, 2019]{BC19}
Basellini, U. and Camarda, C.~G. (2019).
\newblock Modelling and forecasting adult age-at-death distributions.
\newblock {\em Population Studies}, 73(1):119--138.

\bibitem[Basellini et~al., 2023]{BCB22}
Basellini, U., Camarda, C.~G., and Booth, H. (2023).
\newblock {Thirty years on: A review of the Lee-Carter method for forecasting
  mortality}.
\newblock {\em International Journal of Forecasting}, 39(3):1033--1049.

\bibitem[Basellini et~al., 2020]{BKC20}
Basellini, U., Kjaergaard, S., and Camarda, C.~G. (2020).
\newblock An age-at-death distribution approach to forecast cohort mortality.
\newblock {\em Insurance: Mathematics and Economics}, 91:129--143.

\bibitem[Baxter, 1995]{Baxter95}
Baxter, M.~J. (1995).
\newblock Standardization and transformation in principal component analysis,
  with applications to archaeometry.
\newblock {\em Journal of the Royal Statistical Society: Series C},
  44(4):513--527.

\bibitem[Baxter, 2001]{Baxter01}
Baxter, M.~J. (2001).
\newblock Statistical modelling of artefact compositional data.
\newblock {\em Archaeometry}, 43(1):131--147.

\bibitem[Baxter et~al., 2005]{BBC+05}
Baxter, M.~J., Beardah, C.~C., Cool, H. E.~M., and Jackson, C.~M. (2005).
\newblock Compositional data analysis of some alkaline glasses.
\newblock {\em Mathematical Geology}, 37(2):183--196.

\bibitem[{Bergeron-Boucher} et~al., 2017]{BCO+17}
{Bergeron-Boucher}, M.-P., {Canudas-Romo}, V., Oeppen, J., and Vaupel, J.~W.
  (2017).
\newblock Coherent forecasts of mortality with compositional data analysis.
\newblock {\em Demographic Research}, 37:527--566.

\bibitem[Booth, 2006]{Booth06}
Booth, H. (2006).
\newblock Demographic forecasting: 1980 to 2005 in review.
\newblock {\em International Journal of Forecasting}, 22:547--581.

\bibitem[Booth and Tickle, 2008]{BT08}
Booth, H. and Tickle, L. (2008).
\newblock {Mortality modelling and forecasting: A review of methods}.
\newblock {\em Annals of Actuarial Science}, 3(1-2):3--43.

\bibitem[Brouhns et~al., 2002]{BDV02}
Brouhns, N., Denuit, M., and Vermunt, J.~K. (2002).
\newblock Measuring the longevity risk in mortality projections.
\newblock {\em Bulletin of the Swiss Association of Actuaries}, 2:105--130.

\bibitem[Cairns et~al., 2008]{CBD08}
Cairns, A. J.~G., Blake, D., and Dowd, K. (2008).
\newblock {Modeling and management of mortality risk: A review}.
\newblock {\em Scandinavian Actuarial Journal}, 2-3:79--113.

\bibitem[{Canudas-Romo}, 2010]{CanudasRomo10}
{Canudas-Romo}, V. (2010).
\newblock {Three measures of longevity: Time trends and record values}.
\newblock {\em Demography}, 47(2):299--312.

\bibitem[Dellaportas et~al., 2001]{DSS01}
Dellaportas, P., Smith, A.~F., and Stavropoulos, P. (2001).
\newblock Bayesian analysis of mortality data.
\newblock {\em Journal of the Royal Statistical Society: Series A},
  164(2):275--291.

\bibitem[Denuit et~al., 2007]{DDG07}
Denuit, M., Devolder, P., and Goderniaux, A.-C. (2007).
\newblock {Securitization of longevity risk: Pricing survivor bonds with Wang
  transform in the Lee-Carter framework}.
\newblock {\em The Journal of Risk and Insurance}, 74(1):87--113.

\bibitem[Dryden and Mardia, 1998]{DM98}
Dryden, I.~L. and Mardia, K.~V. (1998).
\newblock {\em Statistical Shape Analysis}.
\newblock John Wiley, Chichester.

\bibitem[Egozcue et~al., 2003]{EPM+03}
Egozcue, J.~J., {Pawlowsky-Glahn}, V., {Mateu-Figueras}, G., and
  {Barcel\'{o}-Vidal}, C. (2003).
\newblock Isometric logratio transformations for compositional data analysis.
\newblock {\em Mathematical Geology}, 35:279--300.

\bibitem[Fry et~al., 2000]{FFM00}
Fry, J.~M., Fry, T. R.~L., and McLaren, K.~R. (2000).
\newblock Compositional data analysis and zeros in micro data.
\newblock {\em Applied Economics}, 32(8):953--959.

\bibitem[Gneiting and Raftery, 2007]{GR07}
Gneiting, T. and Raftery, A.~E. (2007).
\newblock Strictly proper scoring rules, prediction and estimation.
\newblock {\em Journal of the American Statistical Association: Review
  Article}, 102(477):359--378.

\bibitem[Graziani and Nigri, 2023]{GN23}
Graziani, R. and Nigri, A. (2023).
\newblock {An age-period-cohort model in a Dirichlet framework: A coherent
  causes of death estimation}.
\newblock Working paper, {SocArXiv}.
\newblock URL: \url{https://osf.io/preprints/socarxiv/856yw}.

\bibitem[Greenacre, 2021]{Greenacre21}
Greenacre, M. (2021).
\newblock Compositional data analysis.
\newblock {\em Annual Review of Statistics and its Application}, 8:271--299.

\bibitem[Greenacre and Grunsky, 2019]{GG19}
Greenacre, M. and Grunsky, E. (2019).
\newblock {The isometric logratio transformation in compositional data
  analysis: A practical evaluation}.
\newblock Working paper 1627, Universitat Pompeu Fabra.
\newblock URL: \url{https://econ-papers.upf.edu/papers/1627.pdf}.

\bibitem[Heligman and Pollard, 1980]{HP80}
Heligman, L. and Pollard, J. (1980).
\newblock The age pattern of mortality.
\newblock {\em Journal of the Institute of Actuaries}, 107(1):49--80.

\bibitem[{Human Mortality Database}, 2023]{HMD2023}
{Human Mortality Database} (2023).
\newblock {\em University of California, Berkeley (USA), and Max Planck
  Institute for Demographic Research (Germany)}.
\newblock Accessed at October 5, 2023. URL: \url{http://www.mortality.org}.

\bibitem[Hurvich and Tsai, 1993]{HT93}
Hurvich, C.~M. and Tsai, C.-L. (1993).
\newblock A corrected akaike information criterion for vector autoregressive
  model selection.
\newblock {\em Journal of Time Series Analysis}, 14(3):271--279.

\bibitem[Hyndman, 2023]{Hyndman23}
Hyndman, R. (2023).
\newblock {\em demography: Forecasting Mortality, Fertility, Migration and
  Population Data}.
\newblock R package version 2.0.

\bibitem[Hyndman et~al., 2013]{HBY13}
Hyndman, R.~J., Booth, H., and Yasmeen, F. (2013).
\newblock Coherent mortality forecasting: the product-ratio method with
  functional time series models.
\newblock {\em Demography}, 50(1):261--283.

\bibitem[Hyndman and Khandakar, 2008]{HK08}
Hyndman, R.~J. and Khandakar, Y. (2008).
\newblock Automatic time series forecasting: the forecast package for {R}.
\newblock {\em Journal of Statistical Software}, 27(3).

\bibitem[Hyndman and Shang, 2009]{HS09}
Hyndman, R.~J. and Shang, H.~L. (2009).
\newblock Forecasting functional time series (with discussion).
\newblock {\em Journal of the Korean Statistical Society}, 38(3):199--221.

\bibitem[Hyndman et~al., 2021]{HZS21}
Hyndman, R.~J., Zeng, Y., and Shang, H.~L. (2021).
\newblock Foreacsting the old-age dependency ratio to determine a sustainable
  pension age.
\newblock {\em Australian and New Zealand Journal of Statistics},
  63(2):241--256.

\bibitem[Kokoszka et~al., 2019]{KMP+19}
Kokoszka, P., Miao, H., Petersen, A., and Shang, H.~L. (2019).
\newblock Forecasting of density functions with an application to
  cross-sectional and intraday returns.
\newblock {\em International Journal of Forecasting}, 35(4):1304--1317.

\bibitem[Kullback and Leibler, 1951]{KL51}
Kullback, S. and Leibler, R.~A. (1951).
\newblock On information and sufficiency.
\newblock {\em The Annals of Mathematical Statistics}, 22(1):79--86.

\bibitem[Lancaster, 1965]{Lancaster65}
Lancaster, H.~O. (1965).
\newblock {The Helmert matrices}.
\newblock {\em The American Mathematical Monthly}, 72(1):4--12.

\bibitem[Lee and Carter, 1992]{LC92}
Lee, R.~D. and Carter, L. (1992).
\newblock {Modeling and forecasting the time series of U.S. mortality}.
\newblock {\em {Journal of the American Statistical Association: Applications
  \& Case Studies}}, 87(419):659--671.

\bibitem[Li et~al., 2020]{LRS20}
Li, D., Robinson, P.~M., and Shang, H.~L. (2020).
\newblock Long-range dependent curve time series.
\newblock {\em Journal of the American Statistical Association: Theory and
  Methods}, 115(530):957--971.

\bibitem[{Mart\'{i}n-Fern\'{a}ndez} et~al., 2003]{MBP03}
{Mart\'{i}n-Fern\'{a}ndez}, J.~A., {Barcel\'{o}-Vidal}, C., and
  {Pawlowsky-Glahn}, V. (2003).
\newblock Dealing with zeros and missing values in compositional data sets
  using nonparametric imputation.
\newblock {\em Mathematical Geology}, 35(3):253--278.

\bibitem[Mazzuco et~al., 2018]{MSZ18}
Mazzuco, S., Scarpa, B., and Zanotto, L. (2018).
\newblock A mortality model based on a mixture distribution function.
\newblock {\em Population Studies}, 72(2):1--10.

\bibitem[Oeppen, 2008]{Oeppen08}
Oeppen, J. (2008).
\newblock {Coherent forecasting of multiple-decrement life tables: A test using
  Japanese cause of death data}.
\newblock In {\em European Population Conference}. European Association for
  Population Studies.
\newblock URL: \url{http://hdl.handle.net/10256/742}.

\bibitem[Pascariu, 2024]{Pascariu24}
Pascariu, M.~D. (2024).
\newblock {\em MortalityLaws: Parametric Mortality Models, Life Tables and
  HMD}.
\newblock R package version 2.1.0.

\bibitem[Pascariu et~al., 2019]{PLC19}
Pascariu, M.~D., Lenart, A., and {Canudas-Romo}, V. (2019).
\newblock The maximum entropy mortality model: Forecasting mortality using
  statistical moments.
\newblock {\em Scandinavian Actuarial Journal}, 2019(8):661--685.

\bibitem[Pollard, 1987]{Pollard87}
Pollard, J.~H. (1987).
\newblock Projection of age-specific mortality rates.
\newblock In {\em Population Bulletin of the United Nations}, 21/22, pages
  55--69. Population Commission.
\newblock URL: \url{https://digitallibrary.un.org/record/42274?ln=en}.

\bibitem[Robine, 2001]{Robine01}
Robine, J.-M. (2001).
\newblock {Redefining the stages of the epidemiological transition by a study
  of the dispersion of life spans: the case of France}.
\newblock {\em Population: An English Selection}, 13(1):173--193.

\bibitem[Scealy et~al., 2015]{SDG+15}
Scealy, J.~L., {de Caritat}, P., Grunsky, E.~C., Tsagris, M.~T., and Welsh,
  A.~H. (2015).
\newblock Robust principal component analysis for power transformed
  compositional data.
\newblock {\em Journal of the American Statistical Association: Theory and
  Methods}, 110(509):136--148.

\bibitem[Scealy and Welsh, 2017]{SW17}
Scealy, J.~L. and Welsh, A.~H. (2017).
\newblock A directional mixed effects model for compositional expenditure data.
\newblock {\em {Journal of the American Statistical Association: Applications
  and Case Studies}}, 112(517):24--36.

\bibitem[Shang et~al., 2011]{SBH11}
Shang, H.~L., Booth, H., and Hyndman, R.~J. (2011).
\newblock {Point and interval forecasts of mortality rates and life expectancy:
  A comparison of ten principal component methods}.
\newblock {\em Demographic Research}, 25(5):173--214.

\bibitem[Shang and Haberman, 2020]{SH20}
Shang, H.~L. and Haberman, S. (2020).
\newblock {Forecasting age distribution of death counts: An application to
  annuity pricing}.
\newblock {\em Annals of Actuarial Science}, 14(1):150--169.

\bibitem[Shang et~al., 2022]{SHX22}
Shang, H.~L., Haberman, S., and Xu, R. (2022).
\newblock Multi-population modelling and forecasting life-table death counts.
\newblock {\em Insurance: Mathematics and Economics}, 106:239--253.

\bibitem[Shang et~al., 2016]{SSB+16}
Shang, H.~L., Smith, P. W.~F., Bijak, J., and Wi\'{s}niowski, A. (2016).
\newblock A multilevel functional data method for forecasting population, with
  an application to the {United Kingdom}.
\newblock {\em International Journal of Forecasting}, 32(3):629--649.

\bibitem[Shannon, 1948]{Shannon48}
Shannon, C.~E. (1948).
\newblock A mathematical theory of communication.
\newblock {\em Bell Labs Technical Journal}, 27(3):379--423.

\bibitem[Tang et~al., 2022]{TWY+22}
Tang, M.-L., Wu, Q., Yang, S., and Tian, G.-L. (2022).
\newblock Dirichlet composition distribution for compositional data with zero
  components: An application to fluorescence in situ hybridization (fish)
  detection of chromosome.
\newblock {\em Biometrical Journal}, 64(4):714--732.

\bibitem[Tsagris et~al., 2016]{TPW16}
Tsagris, M., Preston, S., and Wood, A. T.~A. (2016).
\newblock Improved classification for compositional data using the
  $alpha$-transformation.
\newblock {\em Journal of Classification}, 33:243--261.

\bibitem[Tsagris and Stewart, 2018]{TS18}
Tsagris, M. and Stewart, C. (2018).
\newblock A {D}irichlet regression model for compositional data with zeros.
\newblock {\em Lobachevskii Journal of Mathematics}, 39:398--412.

\bibitem[Tsagris and Stewart, 2020]{TS20}
Tsagris, M. and Stewart, C. (2020).
\newblock A folded model for compositional data analysis.
\newblock {\em Australian \& New Zealand Journal of Statistics},
  62(2):249--277.

\bibitem[{van Raalte} and Caswell, 2013]{VC13}
{van Raalte}, A.~A. and Caswell, H. (2013).
\newblock Perturbation analysis of indices of lifespan variability.
\newblock {\em Demography}, 50(5):1615--1640.

\bibitem[{van Raalte} et~al., 2014]{VMM14}
{van Raalte}, A.~A., Martikainen, P., and Myrskyl\"{a}, M. (2014).
\newblock {Lifespan variation by occupational class: Compression or stagnation
  over time?}
\newblock {\em Demography}, 51:73--95.

\bibitem[Vaupel et~al., 2011]{VZV11}
Vaupel, J.~W., Zhang, Z., and {van Raalte}, A.~A. (2011).
\newblock {Life expectancy and disparity: An international comparison of life
  table data}.
\newblock {\em BMJ Open}, 1(1):e000128.

\bibitem[Wilmoth and Horiuchi, 1999]{WH99}
Wilmoth, J.~R. and Horiuchi, S. (1999).
\newblock {Rectangularization revisited: Variability of age at death within
  human populations}.
\newblock {\em Demography}, 36(4):475--495.

\end{thebibliography}

\end{document}